\documentstyle[nato,psfig]{crckapb} 

\begin{opening}
\title{Full Counting Statistics 
  of Superconductor--Normal-Metal Heterostructures}
\author{W. Belzig}
\institute{Department of Physics, University of Basel,
  Klingelbergstr. 82, 4056 Basel, Switzerland}

\runningtitle{Full Counting Statistics in SN-Structures}
\end{opening}

\begin{document}

\section{Introduction}
\label{cha:intro}

In 1918 Schottky discovered that the fluctuations in vacuum diodes can be
related to the discrete nature of the charge carriers \cite{schottky}. His
observation was that the power spectrum of the current fluctuations gave
direct access to the charge $e$ of the discrete carriers responsible for the
current.  From his theoretical considerations he found a relation between the
noise power of the current fluctuations $S_I$ and the average current $I$,
\begin{equation}
  \label{eq:schottky}
  S_I=2eI\,,
\end{equation}
a result nowadays known as the Schottky formula. Its consequence is, that
the current noise provides information on the transport process, which
is not accessible through conductance measurements only.

Studies of the noise properties of tunnel-junctions renewed the interest
in noise later \cite{schoen:85}. Since about ten years the
investigation of transport in quantum coherent structures has boosted
the interest in the theory of current noise in mesoscopic structures
\cite{blanter,dejong:97}. Correlations in the transport of fermions have
lead to a number of interesting predictions.  For example, the noise of
a single channel quantum contact of transparency $T$ at zero temperature
has the form $S_I=2eI(1-T)$ \cite{khlus,lesovik:89}. The noise is thus
suppressed below the Schottky value, Eq.~(\ref{eq:schottky}). The
suppression is a direct consequence of the Pauli principle, and is
therefore specific to electrons. Particles with bosonic statistics or
classical particles show a different behaviour, e.~g. for Bosons the
noise is enhanced by a factor $1+T$.  A convenient measure of the
deviation from the Schottky result is the so-called Fano factor
$F=S_I/2eI$. For a number of generic conductors, it turns out that the
suppression of the Fano factor is universal, i.~e. it does not depend on
details of the conductor like geometry or impurity concentration. A
diffusive metal with purely elastic scattering leads to $F_{diff}=1/3$
\cite{been-buett:92,nagaev:92}, which is independent on the concrete
shape of the conductor \cite{yuli:94-diffusive}. In a symmetric double
tunnel junction, on the other hand, $F_{dbltun}=1/2$ \cite{chen:91}. A
chaotic cavity (a small region with classical chaotic dynamics,
connected to two leads by open quantum point contacts) shows a
suppression of $F_{cavity}=1/4$ \cite{jalabert:94}. Thus, we conclude
that from a noise measurement two kinds of information can be obtained.
First, we can get information on the statistics of the carriers, e.~g.
if they are fermions.  Second, provided the statistics is known, the
comparison of the magnitude of the noise power with the average current
gives information on the internal structure of the system. However, the
picture described here is a little bit too simplified. In many real
experiments the structure is much less well defined, the temperature is
finite, or other complications make the interpretation of experimental
data less trivial.

Having motivated the interest in the noise, we may ask, what we can
learn further from current fluctuations. \emph{Higher correlators} of
the current will provide even more information on the transport
process.  However, theoretical calculations of higher correlators
become increasingly cumbersome and one should find a different concept
to obtain this information. This step was performed by a transfer of
concepts from the field of quantum optics. Here it is possible to
count experimentally the number of photons occupying a certain quantum
state.  This number is subject to quantum and thermal fluctuations and
requires a statistical description: the \emph{full counting
  statistics} (FCS).  Lesovik and Levitov adopted this terminology to
mesoscopic electron transport
\cite{levitov:93-fcs,levitov:96-coherent}, in which the electrons
passing a certain conductor are counted. Since then the FCS has been
studied in the field of mesoscopic electron transport. It the aim of
this article to review some of the progress, which has been recently
made.

The general problem of counting statistics has been considered before on
a heuristic level. If one makes the \emph{ad hoc} assumption that
individual transfers of charges are \emph{uncorrelated} and
unidirectional, simple calculation show that the probability
distribution of the number of transfered charges is Poissonian
\begin{equation}
  \label{eq:poisson-intro}
  P_{t_0}(N)=e^{-\bar N(t_0)}\frac{{\bar N(t_0)}^N}{N!}\,.
\end{equation}
Here $t_0$ is the time period during which the charges are counted and
$\bar N(t_0)=It_0/e$ is the average number of transfered charges.
Schottky's result (\ref{eq:schottky}) for the current noise power can be
easily derived from Eq.~(\ref{eq:poisson-intro}). This kind of counting
statistics is usually found in tunnel junctions, where the charge
transfers are rare events, or at high temperatures, when the mean
occupation of the states is small and the statistics of the particles
plays no role.  However, in a degenerate electron gas one encounters a
completely different situation: all states are filled due to Fermi
correlations. If we now consider a quantum transport channel of
transmission probability $T$ and applied bias voltage $V$, the rigidity
of the Fermi sea leads to a \emph{fixed} number of electrons, which are
sent into each quantum channel. A charge is transfered to the other side
with probability $T$ and the statistics is therefore binomial
\begin{equation}
  \label{eq:binom-intro}
  P_{t_0}(N)={M(t_0)\choose N} T^N (1-T)^{M(t_0)-N}\,.
\end{equation}
This is the result of the quantum calculation of Levitov and coworkers
\cite{levitov:93-fcs}. Note, that arguments based on the FCS have been
used already to interpret current noise calculations or measurements.
However, in general these interpretations are not unique, and a
calculation of the FCS is required to interpret the results for the
noise unambiguously. Thus, obtaining the FCS is a theoretical task,
which leads to a better understanding of quantum transport processes.

The structure of this article is as follows. In the next section we
introduce the theoretical basis, necessary to obtain the FCS.  Our
approach is based on an extension of the well-known Keldysh-Green's
function technique \cite{yuli:99-annals} and described in Section
\ref{sec:fcs-theory}. A convenient simplification is the \emph{circuit
  theory of mesoscopic transport}, developed by Nazarov
\cite{yuli:94-circuit}, which allows to obtain the FCS for a large
variety of multi-terminal structures with minimal calculational
overhead.  In the two following sections, we will discuss several
concrete examples. First, we show that for phase coherent two-terminal
conductors the counting statistics can be obtained in quite a general
form \cite{belzig:01-super}.  This is illustrated for normal and
superconducting constrictions as well as two-barrier structures.  A
somewhat special case is the diffusive conductor coupled to a
superconductor (i.~e. in the presence of proximity effect), where
intrinsic decoherence between electrons and holes influences the
transport properties \cite{belzig:01-diff}.  Second, we turn to more
complex mesoscopic structures with more than two terminals. An analytic
solution is obtained for the case of an arbitrary number of terminal
connected by tunnel junctions to one central node \cite{boerlin-prl}. As
specific example we calculate the counting statistics of a beam
splitter, in which a normal current or a supercurrent is divided into
two normal currents.  Afterwards we present some conclusions. In the
appendices we give some theoretical background information to the
methods used in this article.

In this article we concentrate on counting statistics in two- or
multi-terminal devices with either normal- or superconducting
contacts.  More aspects of FCS are in other parts of this book.  We
would like to mention works related to FCS not covered here.
Normal-superconductor transport at finite energies and magnetic fields
was recently addressed in \cite{samuelsson1,samuelsson2}.
Time-dependent transport phenomena in normal contacts have been
studied in \cite{levitov:96-coherent}.  Fluctuations of the current in
adiabatic quantum pumps have attracted some attention (see e.g.
\cite{pumps}).  A connection to photon counting or photon transport
can be found in Refs.~\cite{schomerus,kindermann:01-2}.  Counting
statistics has also been addressed in the context of the readout
process of a qubit \cite{makhlin} or to study spin coherence effects
\cite{engel}.  Results for the FCS of entangled electron pairs
\cite{taddei} and of resonant Cooper pair tunneling \cite{choi} have
also been published.

\section{Counting Statistics and Green's function}
\label{cha:theory}

\subsection{Basics of Current Statistics}
\label{sec:fcs}

We introduce some basic formulas, relevant for the theory of FCS.  The
quantity we are after is the probability $P_{t_0}(N)$, that $N$ charges
are transfered in the time interval $t_0$. Equivalently, we can find the
\textit{cumulant generating function} (CGF) $S(\chi)$, defined by
\begin{equation}
\exp(-S(\chi))=\sum_N P_{t_0}(N)\exp(iN\chi)\;. 
\label{eq:cgf-def}
\end{equation}
To keep notations simple in this section, we will limit the discussion
to the two-terminal case, in which only the number $N$ of transfered
charges in one terminal matters. In the other terminal it is given by
$-N$, since the total number of charges is conserved. However, most
relations are straightforwardly generalized to many terminals.  Note
that normalization of the FCS requires that $S(0)=0$.  Also, we will
suppress the explicit dependence of $S(\chi)$ on the
measuring time $t_0$. In the static case considered mostly in this
article, we have $S(\chi)\sim t_0$.

From the full counting statistics one obtains the cumulants
\begin{eqnarray}
  \label{eq:cumulants}
  C_1  =  \overline N\equiv \sum_N N P_{t_0}(N) & , &
  C_2 = \overline{(N-\overline{N})^2}\,, \\
  C_3 = \overline{(N-\overline N)^3} &,&
  C_4 = \overline{(N-\overline N)^4} - 3 \overline{(N-\overline{N})^2}^2\,,
\end{eqnarray}
and so on. The meanings of the various cumulants are depicted in
Fig.~\ref{fig:system}a. Alternatively, the cumulants can be obtained from
the CGF
\begin{equation}
  \label{eq:cumulants1}
  C_n=-\left.
    \left(-i\right)^n\frac{\partial^n}{\partial \chi^n}
    S(\chi)\right|_{\chi=0}\;.
\end{equation}
The relation to the average current and the noise power of current
fluctuations is obtained as follows. Almost trivially one has
\begin{equation}
  \label{eq:average-current}
  C_1=\langle N\rangle = \frac{1}{-e}\int_0^{t_0} dt \langle I(t)\rangle
  =  -\frac{t_0}{e} \bar I\,,
\end{equation}
where we have denoted the charge of the electrons with $-e$. The
relation between the second cumulant and the current noise power
\begin{equation}
  \label{eq:noise-power2}
  S_I\equiv\int_{-\infty}^{\infty} d\tau \left\langle\left\{\delta
  I(\tau),\delta I(0)\right\}\right\rangle 
\end{equation}
is less obvious. We write for the second cumulant
\begin{equation}
  \label{eq:noise-power}
  C_2=\overline{(N-\overline N)^2}=\frac{1}{2e^2}\int_0^{t_0}\int_0^{t_0} dtdt'
  \langle \{\delta I(t), \delta I(t')\}\rangle\,,
\end{equation}
where $\delta I(t)=I(t)-\langle I \rangle $ is the current fluctuation
operator and $\langle ...\rangle$ denote the quantum statistical
average.  We transform the time coordinates to average
$\overline{t}=(t+t')/2$ and relative time $\tau=t-t'$. Assuming the
observation $t_0$ is much larger than the correlation time of the
currents, the correlator in Eq.~(\ref{eq:noise-power}) does not depend on
$T$, and we find the desired relation between current noise power and
the second cumulant
\begin{equation}
  \label{eq:noise-power3}
  S_I=  \frac{2e^2}{t_0}
  \left.\frac{\partial^2 S(\chi)}{\partial \chi^2}\right|_{\chi=0}\,.
\end{equation}
For higher correlators similar formulas can be derived. 

\subsection{Extended Keldysh Green's-function Technique}
\label{sec:fcs-theory}

The task to measure the number of charges transfered in a quantum
transport process has, in general, to be formulated as a quantum
measurement problem. A thorough derivation goes beyond the scope of
this article and we refer to Ref.~\cite{kindermann:01}.  The
quantum-mechanical form of the cumulant generating function is given
by \cite{yuli:99-annals,belzig:01-diff,kindermann:01}
\begin{equation}
  \label{eq:cgf-general}
  e^{-S_(\chi)} =
  \langle {\cal T} e^{-i\frac{\chi}{2e}\int_0^{t_0}dt I(t)}
  \tilde{\cal T} e^{-i\frac{\chi}{2e}\int_0^{t_0}dt I(t)}
  \rangle\,.
\end{equation}
Here ${\cal T}(\tilde{\cal T})$ denotes (anti-)time ordering and $\hat
I(t)$ the operator of the current through a certain cross section.  As
preliminary justification of Eq.~(\ref{eq:cgf-general}) we note, that it
is easily shown that expansion of Eq.~(\ref{eq:cgf-general}) in $\chi$
yields the various current-correlators. The expectation value in
Eq.~(\ref{eq:cgf-general}) can be implemented on the Keldysh contour,
see Appendix \ref{sec:keldysh-green}, which makes it possible to use
diagrammatic methods \cite{rammer}.  Equation (\ref{eq:cgf-general}) has
a form similar to the thermodynamic potential in an external field. From
the linked cluster theorem (see Appendix \ref{sec:cluster}) it follows
that the CGF is the sum of all connected diagrams.  

\begin{figure}[tbp]
  \centerline{\psfig{figure=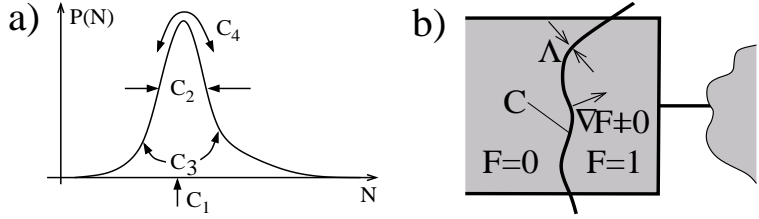,width=0.8\textwidth}}
  \caption{Left panel: An example of a probability distribution 
    illustrating the meaning of the different cumulants. The average
    is given by $C_1$, the width by $C_2$, the 'skewness' by $C_3$ and
    the 'sharpness' is related to the $C_4$. Right panel: schematics of
    an ideal charge counter in a terminal. The number of charges
    passing the cross section $C$ are counted.}
  \label{fig:system}
\end{figure}
To connect the CGF to accessible field-theoretical quantities we
consider the nonlinear response of our electronic circuit to the
time-dependent perturbation
\begin{equation}
  \label{eq:hpert}
  H_{c}(t) = \frac{\chi}{2e} I_c(t) = \mp \frac{\chi}{2e}
  \int d^3 x\Psi^\dagger({\mathbf x},t) \hat j_c({\mathbf x})
  \Psi({\mathbf x},t)\,,
\end{equation}
where the $+(-)$ sign is taken on the lower(upper) part of the Keldysh
contour.  The operator $\hat j_c({\mathbf x})$ is the operator of the
current through a cross section $c$, depicted in Fig.~\ref{fig:system}.
We allow here for multi-component field operators, such as spin or Nambu
for example. Matrices in this subspace are denoted with a $\hat{}$.
Since we are aiming at the total charge counting statistics, we will
assume that $\chi$ is nonzero and constant in a finite time interval
$[0,t_0]$.

The unperturbed system evolves according to a Hamiltonian $H_0=\int d^3x
\Psi^\dagger({\mathbf x}) \hat h_0({\mathbf x}) \Psi({\mathbf x})$,
where $\hat h_0$ is the usual single-particle Hamiltonian of the system.
The equation of motion for the matrix Green's function subject to
$H=H_0+H_c(t)$ reads (in the Keldysh matrix representation)
\begin{equation}
  \label{eq:eom}
  \left[i\frac{\partial}{\partial t} - \hat h_0({\mathbf x}) +
    \frac{\chi}{2e} \bar\tau_3 \hat j_c({\mathbf x})\right] 
  \check G({\mathbf x},t;{\mathbf x'},t';\chi) 
  =\delta(t-t^\prime)\delta({\mathbf x}-{\mathbf x'})\,,
\end{equation}
Here $\bar\tau_3$ denotes the third Pauli matrix in the Keldysh space.
The relation of the Green's function (\ref{eq:eom}) to the CGF
(\ref{eq:cgf-general}) is obtained from a diagrammatic expansion in
$\chi$.  One finds the simple relation (see App.~\ref{sec:cluster})
\begin{equation}
  \label{eq:chi-current}
  \frac{\partial S(\chi)}{\partial \chi} = \frac{it_0}{e} I(\chi)\,.
\end{equation}
The \emph{counting current} $I(\chi)$ is obtained from the
$\chi$-dependent Green's function via
\begin{equation}
  \label{eq:current-general}
  I(\chi)= \int d^3 x \left.
    \textrm{Tr}\left[\bar\tau_3\hat j_c({\mathbf x})
      \check G({\mathbf x},t;{\mathbf x'},t;\chi)\right]
  \right|_{{\mathbf x}\to{\mathbf x'}}\,.
\end{equation}
Since we are assuming a static situation, the r.h.s. of
Eq.~(\ref{eq:current-general}) is time-independent.  The relations
(\ref{eq:eom})-(\ref{eq:current-general}) offer a very
general way to obtain the full counting statistics of any system. It is,
however, difficult to find the Green's function in the general case.

For a mesoscopic transport problem there is a particularly simple way to
access the full counting statistics, based on the separation into
terminals (or reservoirs) and an active part, the first providing
boundary conditions, and the second being responsible for the
resistance.  Let us consider the equation of motion for the Green's
function {\em inside} a terminal for the following parameterization of
the current operator in Eq.~(\ref{eq:hpert})
\begin{equation}
  \label{eq:countfield-choice}
  \hat j_c({\mathbf x})=(\nabla F({\mathbf x}))
  \lim_{{\mathbf x}\to{\mathbf x'}}
  \frac{ie}{2m}\left(\nabla_{\mathbf x}-
    \nabla_{{\mathbf x}'}\right)\hat\sigma_3\,.
\end{equation}
$F({\mathbf x})$ is chosen such that it changes from 0 to 1 across a
cross section C, which intersects the terminal, but is of arbitrary
shape, see Fig.~\ref{fig:system}. Here we have added a matrix
$\hat\sigma_3$ to the current operator, which accounts for possible
multicomponent field operators like in the case of superconductivity.
The change from 0 to 1 should occur on a length scale $\Lambda$, for
which we assume $\lambda_F\ll\Lambda\ll l_{imp},\xi_0$ (Fermi wave length
$\lambda_F$, impurity mean free path $l_{imp}$, and coherence length
$\xi_0=v_F/2\Delta$). Under this assumption we can reduce Eq.~(\ref{eq:eom})
to its quasiclassical version (see Ref.~\cite{eilenberger} and
App.~\ref{sec:eilenberger}). This is usually a very good approximation,
since all currents in a real experiment are measured in normal
Fermi-liquid leads.  

The Eilenberger equation in the vicinity of the cross section reads
\begin{equation}
  \label{eq:eilenberger-counting}
  {\mathbf v_F }\nabla \check g({\mathbf{x}},{\mathbf v_F},t,t',\chi)
  = \left[-i\frac{\chi}{2} (\nabla F({\mathbf x})) {\mathbf v_F} \check \tau_K 
  \,,\,\check g({\mathbf{x}},{\mathbf v_F},t,t',\chi) \right] \,.
\end{equation}
Here $\check\tau_K=\bar\tau_3\hat\sigma_3$ is the matrix of the
current operator. Other terms can be neglected due to the assumptions
we have made for $\Lambda$.  The counting field can then be eliminated
by the gauge-like transformation
\begin{equation}
  \check g({\mathbf x},{\mathbf v_F},t,t',\chi) 
  = e^{-i\chi F({\mathbf x})\check \tau_K/2}
  \check g({\mathbf x},{\mathbf v_F},t,t',0) 
  e^{i\chi F({\mathbf x})\check \tau_K/2}\,.
\end{equation}
We assume now that the terminal is a diffusive metal of negligible
resistance. Then the Green's functions are constant in space (except in
the vicinity of the cross section C) and isotropic in momentum space.
Applying the diffusive approximation \cite{usadel} in the terminal
leads to a transformed terminal Green's function
\begin{equation}
  \label{eq:countrot}
  \check G(\chi) = e^{-i\chi\check \tau_K/2}
  \check G(0) e^{i\chi\check \tau_K/2}\,,
\end{equation}
on the right of the cross section $C$ (where $F({\mathbf x}=1$) with
respect to the case without counting field.  Consequently the counting
field is entirely incorporated into a \textit{modified boundary
  condition} imposed by the terminal onto the mesoscopic system. Note,
that it follows from (\ref{eq:countfield-choice}) and
(\ref{eq:eilenberger-counting}), that the counting field for a
particular terminal vanishes from the equations of motion in the
mesoscopic system and the other terminals.

The generalization of this method to the counting statistics for
multiterminal structures was performed in \cite{yuli:01-multi}. The
surprisingly simple result is, that one has to add a separate counting
field for each terminal, in which charges are counted. In
Sec.~\ref{sec:multi-theory} we demonstrate this for an example, in
which an analytic solution can be found.  This concludes the
derivation of our theoretical method to obtain the FCS.

What are the achievements of this method? We should emphasize that it
does not simplify the solution of a specific transport problem, i.e. we
still have to know the solution corresponding to the Hamiltonian $H_0$.
If this solution is not known, the counting field makes this situation
not easier. Rather, the method paves a very general way to obtain the
FCS, if a method to find the average currents, i.e. for $\chi=0$, is
known. In the next section we will introduce such a method, the circuit
theory of mesoscopic transport. Initially it was invented to calculate
average currents, however the method to obtain the FCS introduced
in this section is straightforwardly included.

What is the price to pay? Loosely speaking, the method to obtain
the average currents has to be sufficiently general. Usually the absence
of a field, which has different signs on the upper and lower part of the
Keldysh contour, allows some simplification. For example, in the
Keldysh-matrix representation all Green's functions can be brought into
a tri-diagonal form, which is obviously simpler to handle than the full
matrix. The method above does not allow this simplification anymore. Or,
in other words, the counting rotation (\ref{eq:countrot}) destroys the
triangular form. Thus, the price we have to pay for an easy
determination of the FCS is that we need a method, which respects the
\emph{full Keldysh-matrix structure} in all steps. The circuit theory, which we
describe in the next section fulfills this requirement.

\subsection{Circuit Theory}
\label{sec:circuit}

A concise formulation of mesoscopic transport is the so-called circuit
theory \cite{yuli:94-circuit,yuli:99-supplat}.  Its main idea, borrowed
from Kirchhoff's classical circuit theory, is to represent a mesoscopic
device by discrete elements. These approximate the layout of an
experimental device with arbitrary accuracy, provided one chooses enough
elements. In practice one has to find the balance between a small grid
size and the computational effort.

We briefly repeat the essentials of the circuit theory. Topologically,
one distinguishes three elements: terminals, nodes and connectors.
Terminals are the connections to the external measurement circuit and
provide boundary conditions, specifying externally applied voltages,
currents or phase differences. Besides, they also determine the type of
the terminal, i.e. if it is a normal metal or a superconductor. The
actual circuit, which is to be studied consists of nodes and connectors,
the first determining the approximate layout, and the second describing the
connections between different nodes.

The central element of the circuit theory is the arbitrary connector,
characterized by a set of transmission coefficients $\{T_n\}$. Its
transport properties are described by a {\em matrix current} found in
\cite{yuli:99-supplat}
\begin{equation}
  \label{eq:matrix-current}
  \check{I}_{12}=-\frac{e^2}{\pi}\sum_n 
  \frac{2T_n\left[\check G_1,\check G_2\right]}{
    4+T_n\left(\{\check G_1,\check G_2\}-2\right)}\,.
\end{equation}
Here $\check G_{1(2)}$ denote the matrix Green's functions on the left
and the right of the contact. We should emphasize that the matrix form
of (\ref{eq:matrix-current}) is crucial to obtain the FCS, since it is
valid for any matrix structure of the Green's functions. The
\emph{electrical current} is obtained from the matrix current by
\begin{equation}
  \label{eq:el-current}
  I_{12}=\frac{1}{4e}\int dE {\rm Tr} \check\tau_K\check I_{12}\,.
\end{equation}

A special case is a diffusive wire of length $L$ in the presence of
proximity effect. If $L$ is longer than other characteristic lengths
like $\xi_0$, decoherence between electrons and holes becomes important
and the transmission eigenvalue ensemble is not known. Instead, one
solves the diffusion-like equation \cite{usadel}
\begin{equation}
  \label{eq:usadel}
  \nabla D({\mathbf x})\check G({\mathbf x}) 
  \nabla\check G({\mathbf x}) =
  \left[ -i E\hat\tau_3 , \check G({\mathbf x})\right]\,.
\end{equation}
The matrix current is now given by
\begin{equation}
  \label{eq:matrix-current-diffusive}
  \check I({\mathbf x}) = -\sigma({\mathbf x}) 
  \check G({\mathbf x}) \nabla\check
  G({\mathbf x})\,.
\end{equation}
In these equations, $D({\mathbf x})$ the diffusion coefficient and
$\sigma({\mathbf x})=2e^2N_0D({\mathbf x})$ is the conductivity. In general
this equation can only be solved numerically, but in some special cases
(e.g. for $E=0$) an analytic solution is possible.

If the circuit consists of more than one connector, the transport
properties can be found from the circuit theory by means of the
following circuit rules.  We take the Green's functions of the terminals
as given and introduce for each internal node an (unknown) Green's
function. The two rules determining the transport properties of the
circuit completely are
\begin{enumerate}
\item $\check G^2_j=\check 1$ for the Green's functions of all internal
  nodes $j$. 
\item The total matrix current in a node is conserved: $\sum_i \check
  I_{ij}=0$, where the sum goes over all nodes or terminals connected to
  node $j$ and each matrix current is given by
  (\ref{eq:matrix-current}) or (\ref{eq:matrix-current-diffusive}), depending
  on the type of the connector.
\end{enumerate}

An important feature of the circuit theory in the form presented above
is that it accounts for any matrix structure (i.e. Keldysh, Nambu, Spin,
etc.). Thus, we can straightforwardly obtain the FCS along the lines of
Sec.~\ref{sec:fcs-theory}.  If the charges in a terminal are counted, we
have to apply a counting rotation (\ref{eq:countrot}) to the terminal
Green's function. The counting-rotation matrix has the form
$\check\tau_K=\hat\sigma_3\bar\tau_3$, where $\hat\sigma_i (\bar\tau_i)$
denote Pauli-matrices in Nambu(Keldysh)-space. Then we can employ the
circuit rules to find the $\chi$-dependent Green's function and finally
obtain the total CGF by
integrating all currents into the terminals over their respective
counting fields (see \cite{yuli:01-multi} and \cite{boerlin-prl} for
more details).

\section{Two-Terminal Contacts}
\label{cha:two}

\begin{figure}[t]
    \centerline{\psfig{figure=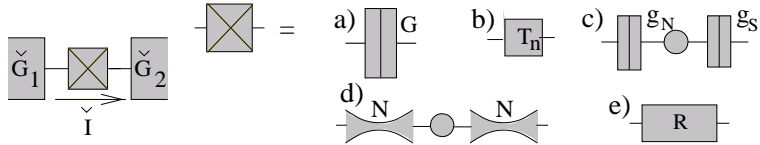,width=0.8\textwidth}}
    \caption{Two terminal contacts. The leftmost picture shows a general
      two-terminal contact. A matrix current $\check I$ is flowing
      between terminals, characterized by Green's functions $\check
      G_{1,2}$. Several connectors (a-e) are described in the text and
      depicted as: a) tunnel junction, b) arbitrary connector with
      transmission eigenvalues $\{T_n\}$, c) two tunnel junctions in
      series with conductances $g_N(g_S)$, respectively, d) a chaotic
      cavity connected by $N_{ch}$ channels to the leads, and e) a
      diffusive wire, characterized by a resistance $R_N$. }
    \label{fig:two-term}
\end{figure}

In this chapter we demonstrate several examples of the FCS of contacts
between two terminals, see Fig.~\ref{fig:two-term}. One can easily
derive a number of rather general results, such as Poisson statistics
in the case of tunnel junction, or binomial statistics for single
channel contacts of transparency $T$. All these results can be found
from a general CGF \cite{belzig:01-super}, which depends on the
ensemble of transmission eigenvalues $\{T_n\}$. To illustrate specific
examples, we will compare the cases of transport between two normal
terminals or between one superconducting and one normal terminal. In
the end, the statistics of an equilibrium supercurrent is discussed.

\subsection{Tunnel Junction}

The counting statistics of a tunnel junction contact can be obtained from a
direct expansion of the matrix current (\ref{eq:matrix-current}), if $T_n\ll 1
$ for all $n$. It coincides with the result obtained by means of the tunneling
Hamiltonian \cite{levitov:tunnel}. The matrix current takes the form
\begin{equation}
  \label{eq:tunnel-current}
  \check I_{\rm tun}(\chi) = \frac{G_T}{2} 
  \left[\check G_1(\chi),\check G_2\right]\,.
\end{equation}
Here the matrix current depends only on the tunneling conductance $G_T =
(e^2/\pi)$ $ \sum_n T_n$ of the contact, and we have arbitrarily chosen to
count the charges in terminal 1. Now, the counting current is
\begin{equation}
  \label{eq:tunnel-current2}
  I(\chi)=\frac{G_T}{8e}\int dE \textrm{Tr}\left(
    \check\tau_K\left[\check G_1(\chi),\check G_2\right]\right)\,.
\end{equation}
We use the pseudo-unitarity $\check\tau_K^2=\check 1$ to express the
counting rotation as
$\exp(i\chi\check\tau_K/2)=\check\tau_+\exp(i\chi/2)+\check\tau_-\exp(-i\chi/2)$,
where $\check \tau_{\pm}=(1\pm\check\tau_K)/2$.  Then the current has
the form
\begin{eqnarray}
  \label{eq:tunnel-current3}
  I(\chi) & = & \frac{e}{t_0}\left[ N_{12}e^{-i\chi} -N_{21}e^{i\chi}\right]\,,\,\\
  N_{ij} & = & \frac{t_0G_T}{4e^2}\int dE \textrm{Tr}\left[\check\tau_+\check
  G_i\check\tau_-\check G_j\right]\,.
\end{eqnarray}
The CGF follows from integrating (\ref{eq:tunnel-current3}) with respect
to $\chi$ and we obtain
\begin{equation}
  \label{eq:cgf-tunnel}
  S(\chi) = -N_{12}(e^{i\chi}-1)-N_{21}(e^{-i\chi}-1)\,.
\end{equation}
It is easy to see that the even and odd cumulants obey
\begin{equation}
  \label{eq:tunnel-cumulants}
  C_{2n+1}=N_{12}-N_{21}\quad,\quad C_{2n}=N_{12}+N_{21}\,.
\end{equation}
If only tunneling processes in one direction occur (say from 1 to 2),
$N_{21}=0$ and the average $\bar N=N_{12}$. The statistics is Poissonian
\begin{equation}
  \label{eq:poisson}
  P_{t_0}(N)=e^{-\bar N}\frac{\bar N^N}{N!}
\end{equation}
In particular, for the current noise we find the Schottky result
$S_{I}=2eI$.

We conclude that the charge counting statistics of a tunnel junction (or
more precisely, if the transfer events are rare) is of a generalized
Poisson type \cite{levitov:tunnel}.  If only tunneling events in one
directions are possible, the statistics is Poissonian.

\subsection{Quantum Contact -- General Connector}
\label{sec:qpc}

We consider now a quantum contact characterized by a set of transmission
eigenvalues $\{T_n\}$. It turns out that the CGF can be obtained in a
quite general form, valid for arbitrary junctions between
superconductors and/or normal metals. 

The matrix current through a quantum contact is described by
Eq.~(\ref{eq:matrix-current}) and the CGF can then be found from the
relation $\partial S(\chi)/\partial\chi=(-it_0/4e^2)$ $\int dE$ $
{\rm{Tr}}[\check{\tau}_{\rm K}\check I(\chi)]$.  Using that $[\check
A,\{\check A,\check G_2\}]=0$ for all matrices with $\check A^2=\check
1$ the following identity holds
\begin{equation}
  \label{eq:matrix-identity}
  \mbox{Tr}\frac{\partial}{\partial\chi} 
  \left\{\check G_1(\chi),\check G_2\right\}^n =
  \frac{i}{2}\mbox{Tr}\check\tau_K\left[\check G_1(\chi),\check G_2\right]
  \left\{\check G_1(\chi),\check G_2\right\}^{n-1}\,.
\end{equation}
We can therefore integrate (\ref{eq:el-current}) with respect to
$\chi$ and obtain \cite{belzig:01-super}
\begin{equation}
  \label{eq:cgf-two-terminal}
  S(\chi)=-\frac{t_0}{2\pi}\sum_n\int dE {\textrm{Tr}}
  \ln\left[1+\frac{T_n}{4}
    \left(\{\check G_1(\chi),\check G_2\}-2\right)\right]\;.
\end{equation}
This is a very important result. It shows that the counting statistics
of a large class of constrictions can be cast in a common form,
independent of the contact types. Another important property of
Eq.~(\ref{eq:cgf-two-terminal}) is, that the CGF's of all constrictions
are linear statistics of the transmission eigenvalue distribution, and
can therefore be averaged over by standard means (see e.g.
\cite{beenakker:97-rmp}). Examples are given in the following sections.
Note also, that an expansion of Eq.~(\ref{eq:cgf-two-terminal}) for $T_n\ll
1$ yields the result (\ref{eq:cgf-tunnel}) of the tunnel case.

We will now discuss several illustrative examples. Consider first two
normal reservoirs with occupation factors $f_{1(2)} =
[\exp((E-\mu_{1(2)})/T_e)+1]^{-1}$ ($T_e$ is the temperature). We
obtain the result \cite{levitov:93-fcs,levitov:96-coherent}
\begin{eqnarray}
  \label{eq:cgf-normalcontact}
  S(\chi)=-\frac{t_0}{\pi}\sum_n\int dE
  \ln\left[1+B_{1n}(E)\left(e^{i\chi}-1\right)
    +B_{-1n}(E)\left(e^{-i\chi}-1\right)\right]\,.
\end{eqnarray}
Here we introduced the probabilities $B_{1n}=T_n
f_1(E)\left(1-f_2(E)\right)$ for a tunneling event from 1 to 2 and
$B_{-1n}$ for the reverse process.  The terms with {\em counting
  factors} $e^{\pm i\chi}-1$ obviously correspond to charge transfers
from 1 to 2 (2 to 1). At zero temperature and $\mu_1-\mu_2=eV\ge 0$ the
integration can easily be evaluated and we obtain
\begin{equation}
  \label{eq:cgf-normal-zero-temp}
  S(\chi)=-\frac{et_0|V|}{\pi}\sum_n
  \ln\left[1+T_n
    \left(e^{i\chi}-1\right)\right]\,.
\end{equation}
The corresponding statistics for a single channel with transparency
$T$ is binomial
\begin{equation}
  \label{eq:binom}
  P_{t_0}(N)={M \choose N} T^{N}(1-T)^{M-N}\,.
\end{equation}
Here we have introduced the \emph{number of attempts} $M=e t_0V/\pi $,
which is the maximal number of electrons that can be sent through one
(spin-degenerate) channel in a time interval $t_0$ due to the exclusion
principle for Fermions.

The FCS of an superconductor-normal metal-contact also follows from
Eq.~(\ref{eq:cgf-two-terminal}) (for a definition of the various Green's
functions see App.~\ref{sec:bulk-solutions}).  Evaluating the trace in
Eq.~(\ref{eq:cgf-two-terminal}) the CGF can be expressed as
\cite{khmelnitzkii}
\begin{equation}
  \label{eq:cgf-sncontact}
   S(\chi)=-\frac{t_0}{2\pi}\sum_n\int dE
  \ln\left[1+\sum\limits_{q=-2}^{2} 
    A_{nq}(E)\left(e^{iq\chi}-1\right)\right]\,.
\end{equation}
The coefficients $A_{nq}(E)$ are related to a charge transfer of
$q\times e$.  For example a term $\exp(2i\chi)-1$ corresponds just to an
Andreev reflection process, in which two charges are transfered
simultaneously \cite{2e}.  Explicit expressions for the various
coefficients are given in Appendix \ref{sec:coefficients-sn}. For the
BCS case, they reproduce the results of Ref.~\cite{khmelnitzkii}. In the
fully gapped single-channel case at energies $k_BT_{e}\ll eV \ll\Delta$
only terms corresponding to Andreev reflection ($A_{2}$) are nonzero and
the CGF becomes
\begin{equation}
  S(\chi)=-\frac{et_0 |V|}{\pi} 
  \ln\left[1+R_A\left(e^{i2\chi}-1\right)\right]\,,
  \label{eq:cgf-andreev}
\end{equation}
where $R_A=T^2/(2-T)^2$ is the probability of Andreev reflection. The
CGF is now $\pi$-periodic, which means that only \emph{even} numbers of
charges are transfered, a consequence of Andreev reflection. The
corresponding statistics is binomial
\begin{equation}
  P_{t_0}(2N)={ M \choose N } R_A^N\left(1-R_A\right)^{M-N}\,,\, 
  P_{t_0}(2N+1)=0\,.
\end{equation}
The number of attempts $M$ is, however, the same as in the normal state.

It is interesting to see how the CGF for normal transport, i.~e.,
Eq.~(\ref{eq:cgf-normalcontact}), emerges from
Eq.~(\ref{eq:cgf-sncontact}).  Putting $f_{R,A}=0$ and $g_{R,A}=\pm 1$
the coefficients in Eq.~(\ref{eq:cgf-sncontact}) can be written as
\begin{eqnarray*}
  A_{\pm 1} & = & B_{\pm 1}^+ + B_{\pm 1}^- -2  B_{\pm 1}^+ 
  B_{\pm 1}^-  - B_{\pm 1}^+ B_{\mp 1}^-- B_{\pm 1}^- B_{\mp 1}^+\,, \\
  A_{\pm 2} & = & B_{\pm 1}^+ B_{\pm 1}^-\;\;;\;\;
  B_1^{\pm} = T B_1(\pm E)\,,
  B_{-1}^{\pm} = T B_{-1}(\pm E)\,,
\end{eqnarray*}
The argument of the 'ln' in Eq.~(\ref{eq:cgf-sncontact}) factorizes in
positive and negative energy contributions
\begin{eqnarray}
  \ln\left[1+\sum\limits_{q=-2}^{2} 
    A_{q}\left(e^{iq\chi}-1\right)\right]
  =\sum\limits_{s=\pm}\ln\left[1+\sum\limits_{q=-1}^{1} 
    B_{q}^{s}\left(e^{iq\chi}-1\right)\right]\,.
\end{eqnarray}
Integrating over energy both terms give the same contribution, and the
CGF results in Eq.~(\ref{eq:cgf-normalcontact}). This shows explicitly
how the positive and negative energy quasiparticles are correlated in
the Andreev reflection process.

\subsection{Double Tunnel Junction}
\label{sec:dbltun}

We now consider a diffusive island (or a chaotic cavity) connected to
two terminals by tunnel junctions with respective conductance $g_1$ and
$g_2$ \cite{boerlin-prl}. We assume for the conductance of the island
$g_{island}\gg g_{1,2}\gg e^2/h$, so we can neglect charging effects.
This provides a simple application of the circuit theory. The layout is
shown in Fig.~\ref{fig:two-term}c. The central node is described by an
unknown Green's function $\check G_c$.  We have two matrix currents
entering the node, which obey a conservation law:
\begin{equation}
  0= \check I_1+\check I_2 = \frac{1}{2} 
  \left[g_1\check G_1+g_2 \check G_2,\check G_c\right]\,.
\end{equation}
Using the normalization condition $\check G_c^2=1$ the solution is
\begin{equation}
  \label{eq:dbltun-solution}
  \check G_c=\frac{g_1\check G_1+g_2 \check G_2}{
    \sqrt{g_1^2+g_2^2 + g_1g_2\left\{\check G_1,\check
    G_2\right\}}}\,. 
\end{equation}
We can integrate the current $I(\chi)\sim \rm{Tr}
\check \tau_K \check I_1$ and obtain the CGF
\begin{equation}
  \label{eq:cgf-dbl-general}
  S(\chi) = -\frac{t_0}{4e^2}\int dE
  {\rm Tr}\sqrt{g_1^2+g_2^2+g_1g_2\left\{\check G_1,\check G_2\right\}}\;.
\end{equation}
Again, this result is valid for all types of contacts between
normal metals and superconductors. 

We first evaluate the trace for two normal leads and find
\begin{eqnarray}
  \label{eq:cgf-dbl-normal}
  \lefteqn{S(\chi) = - \frac{t_0}{2e}\int dE\times} \\\nonumber &
  \sqrt{(g_1+g_2)^2+4g_1g_2 \left(f_1(1-f_2)(e^{i\chi}-1)
      + f_2(1-f_1)(e^{-i\chi}-1\right))}\,.
\end{eqnarray}
We observe that the CGF contains again counting factors corresponding to
charge transfer from 1 to 2 and vice versa. In contrast to the
Poissonian case for a tunnel junction, Eq.~(\ref{eq:cgf-tunnel}), the
charge transfers are not independent, but correlated by the square-root
function. At zero temperature and $\mu_1-\mu_2=eV>0$ we find the result,
\begin{equation}
  S(\chi)=-\frac{t_0 V}{2e}\sqrt{(g_1+g_2)^2+4g_1g_2 (e^{i\chi}-1)}\,.
\end{equation}
There are two relatively simple limits. If the two conductances are very
different (e.g. $g_1\ll g_2$), we return to Poissonian statistics:
\begin{equation}
  \label{eq:dbl-poisson}
  S(\chi)=- \frac{t_0Vg_1}{e} (e^{i\chi}-1)\,.
\end{equation}
On the other hand, in the symmetric case $g_1=g_2=g$ we find \cite{dejong}
\begin{equation}
  \label{eq:cgf-dbl-sym}
  S(\chi)=-\frac{t_0Vg}{e}(e^{i\chi/2}-1)\,,
\end{equation}
and the cumulants are
\begin{equation}
  C_n=\frac{\bar N}{2^{n-1}}\quad,\quad \bar N=\frac{t_0g|V|}{2e}\,.
\end{equation}
The suppression factor $1/2$, which occurs already in the Fano factor,
carries forward to all cumulants. Note, that the same kind of statistics
(\ref{eq:cgf-dbl-normal}) follows also from a Master
equation \cite{dejong,bagrets}.

The CGF (\ref{eq:cgf-dbl-general}) for the transport between a normal
metal and a superconductor (at zero temperature, for simplicity) reads
\cite{boerlin-prl}
\begin{equation}
  \label{eq:cgf-dbl-sncontact}
  S(\chi)=-\frac{t_0V}{e\sqrt 2} 
  \sqrt{g_1^2+g_2^2+\sqrt{\left(g_1^2+g_2^2\right)^2
      +4g_1^2g_2^2(e^{i2\chi}-1)}}\,.
\end{equation}
Thus, the influence of the superconductor is two-fold: charges are
transfered in units of $2e$ (indicated by the $\pi$-periodicity) and
another square root is involved in the CGF, resulting from the higher
order correlations. In the limit that both conductances are very
different (e.g. $g_1\ll g_2$) we obtain again Poissonian statistics
\begin{equation}
  S(\chi)=-\frac{t_0V}{e} \frac{g_1^2}{g_2} 
  \left(e^{i2\chi}-1\right)\,.
\end{equation}
This corresponds to uncorrelated transfers of pairs of
charges. Consequently we obtain for the cumulants
\begin{equation}
  \label{eq:dbl-qeff}
  C_n=2^{n-1} \bar N\quad,\quad \bar N=\frac{2 t_0 V g_1^2}{eg_2}\,,
\end{equation}
and the effective charge $2e$ can indeed be found from the Fano factor.
The transport properties at finite energies and magnetic fields of this
structure have recently been addressed in \cite{samuelsson1} and
\cite{samuelsson2}. 

\subsection{Symmetric Chaotic Cavity}
\label{sec:cavity}

Another interesting system is the chaotic cavity, i.~e. a small island
coupled to terminals by perfectly transmitting contacts (with
$N_{ch}\gg 1$ channels). This system is described by the circuit
depicted in Fig.~\ref{fig:two-term}d. The matrix current between
terminal 1 and the cavity is $\check I_1= N_{ch}(e^2/\pi) [\check
G_1,\check G_c]/(2+\{\check G_1,\check G_c\})$.  Similar as in the
previous chapter, the current conservation reads now
\begin{equation}
  0=\frac{\left[\check G_1,\check G_c\right]}{
    2+\left\{\check G_1,\check G_c\right\}}+
  \frac{\left[\check G_2,\check G_c\right]}{
    2+\left\{\check G_2,\check G_c\right\}}\,,
\end{equation}
which is solved by the solution (\ref{eq:dbltun-solution}) for
$g_1=g_2$.  The integration of (\ref{eq:el-current}) leads to
\begin{equation}
  \label{eq:cgf-cavity}
  S(\chi)=-N_{ch}\frac{t_0}{2\pi}\int dE \textrm{Tr}
  \ln\left[2+\sqrt{2+
      \left\{\check G_1(\chi)\,,\,\check G_2\right\}}\right]\,.
\end{equation}
The interpretation of this result is straightforward. As we have seen
in Sec.~\ref{sec:qpc}, the $\ln$ appeared already in the FCS of a
quantum point contact (leading to binomial statistics). The
square-root we encountered already in the previous section and we
attribute it to inter-mode mixing on the central node ('cavity
noise').

For normal leads at zero temperature and applied bias voltage $V$ we
obtain (with the number of attempts $M=N_{ch}t_0eV/\pi$)
\begin{equation}
  \label{eq:cgf-cavity-normal}
  S(\chi)=-2M\ln\left[1+e^{i\chi/2}\right]\,.
\end{equation}
On the other hand, in the case of Andreev reflection we find
\begin{equation}
  \label{eq:cgf-cavity-andreev}
  S(\chi)=-2M\ln\left[2+e^{i[\chi {\rm mod}
  \pi]}+2\sqrt{1+e^{i[\chi {\rm mod} \pi]}}\right]\,,
\end{equation}
where the $\pi$-periodicity reflects the fact that charges are
transfered in pairs.

\subsection{Diffusive Connector}
\label{sec:diffusive}

A metallic strip of length $L$ with purely elastic scattering,
characterized by an elastic mean free path $l$, is called a diffusive
connector if $l\ll L$. Its transport properties are governed by the
diffusion-like Usadel equation, Eq.~(\ref{eq:usadel}). In the case of
proximity transport the r.h.s. of Eq.~(\ref{eq:usadel}) accounts for
decoherence of electrons and hole during their diffusive motion along
the normal wire.  This term has the form of a \emph{leakage current}, if
the l.h.s. is considered as a conservation law for the matrix current
\cite{yuli:99-supplat}.  Note that the electric current is still
conserved, it is only loss of coherence which occurs. In general the
solution of the full equation is rather complicated and can only be found
numerically for the full parameter range. There is, however, one case,
in which an analytic solution is possible, namely if the r.h.s. of
Eq.~(\ref{eq:usadel}) vanishes. This is either the case for purely
normal transport, when electrons and holes are transported
independently, or for $E=0$, which means we are restricted to low
temperatures and voltages. The scale here is set by the Thouless energy
$E_{Th}$ (given by $\hbar D/L^2$ for a wire of uniform cross section).
At this scale the famous reentrance effect of the conductance occurs
\cite{yuli:96}.  This regime was studied in Ref.  \cite{belzig:01-diff}
and will be discussed in connection with experimental results for the
current noise in the article by Reulet \emph{et al.} in this book.
Numerical results for equilibrium counting statistics in the full
parameter range are discussed below.

We now concentrate on the analytic solution in a quasi-one-dimensional
geometry, i.~e. we assume a wire of uniform cross section connects two
reservoirs located at $x=0$ and $x=L$. It is characterized by a conductivity
$\sigma(x)$, which in general could depend on $x$, e.~g. due to an
inhomogeneous concentration of scattering centers.  The diffusion
equation is then indeed a conservation law for the matrix current
density
\begin{equation}
  \label{eq:usadel-zero-energy}
  \frac{\partial}{\partial x}\check{j}=0\quad,\quad 
  \check j=-\sigma(x)\check G(x)\frac{\partial}{\partial x}
  \check G(x)\,.
\end{equation}
This equation has to be solved with the boundary condition that $\check
G(0)=\check G_1$ and $\check G(L)=\check G_2$. It follows from
Eq.~(\ref{eq:usadel-zero-energy}) and the normalization condition
$\check G^2(x)=\check 1$ that $\check G(x)$ obeys the equation
\begin{equation}
  \label{eq:usadel2}
   \check{j}={\textrm{const.}}\quad,\quad 
   \frac{\partial}{\partial x}\check G(x)=
   -\frac{1}{\sigma(x)}\check G(x)\check{j} \,. 
\end{equation}
This is an homogeneous first-order differential equation, which is easily
solved. Using the boundary conditions we find the solution
\begin{equation}
  \label{eq:usadel2-solution}
  \check G(L)=\check G(0) 
  e^{\check I/g_d}\;,\; g_d=\frac{A}{\int_0^Ldx/\sigma(x)}\,,
\end{equation}
where $g_d$ is the conductance of the wire and $A$ its cross section.
The current is thus given by \cite{yuli:99-annals}
\begin{equation}
  \label{eq:diff-current}
  \check I(\chi) = - g_d \ln\left(\check G_1(\chi)\check G_2\right)\,,
\end{equation}
where we have reinserted the dependence on the counting field $\chi$.
To find the CGF we have to find the integral $\int d\chi$Tr$\check\tau_K\check
I(\chi)$ with respect to the counting field. Expanding the $\ln$ and
using repeatedly the normalization condition the results is
\begin{equation}
  \label{eq:cgf-diffusive}
  S(\chi)=-\frac{t_0 g_d}{8e^2}\int dE \textrm{Tr}\left[
    \textrm{acosh}^2\left(
      \frac12\left\{\check G_1(\chi),\check G_2\right\}\right)\right]\,.
\end{equation}
This is the counting statistics for a general diffusive contact (under
the restrictions mentioned above). The first thing to note is that from
the properties of the wire only the conductance enters and this holds
for all cumulants. In this sense Eq.~(\ref{eq:cgf-diffusive}) shows that
the entire FCS is universal. In our derivation, we have assumed a wire
of uniform cross section, but it has been shown \cite{yuli:94-diffusive}
that this also holds for an arbitrary shape of the wire (as long as it
can be considered as quasi one-dimensional). We should also mention,
that one could have obtained the same result, by averaging the CGF
(\ref{eq:cgf-general}) over the bimodal distribution of transmission
eigenvalues \cite{dorokhov,yuli:94-diffusive}.

Now we evaluate the trace in Eq.~(\ref{eq:cgf-diffusive}) for normal
metals at zero temperature and applied bias voltage $eV$. We obtain
\begin{equation}
  \label{eq:cgf-diff-normal}
  S(\chi)=-\frac{t_0 g_d V}{4e} \textrm{acosh}^2\left(2e^{i\chi}-1\right)\,,
\end{equation}
which coincides with the results of
Refs.~\cite{yuli:99-annals,levitov:96-diffusive}. 

In the case of Andreev transport the easiest way to obtain the CGF is as
follows. We have already previously noted, that
Eq.~(\ref{eq:cgf-diff-normal}) follows from averaging
Eq.~(\ref{eq:cgf-normal-zero-temp}) with the transmission eigenvalue
distribution $ \rho(T)=(2e^2/g_d\pi)/T\sqrt{1-T}$ for a diffusive metal
\cite{yuli:94-diffusive,dorokhov}. Now, the CGF for Andreev transport
Eq.~(\ref{eq:cgf-andreev}) has the same form as in the case of normal
transport, provided we replace $\chi$ with $2\chi$ and the transmission
eigenvalue $T_n$ with the Andreev reflection probability $R_A(T_n)$. A
simple calculation shows, that the $R_A$ are distributed according to
the \emph{same} distribution as the normal transmission eigenvalues (up
to a factor of 1/2). Thus, we can immediatly read off the CGF for the
diffusive SN-wire in the limit of zero temperature and $eV\ll E_{Th}$
from Eq.~(\ref{eq:cgf-diff-normal}) and obtain
\begin{equation}
  \label{eq:cgf-diffusive-andreev}
  S(\chi)=-\frac{t_0 g_d V}{8e} \textrm{acosh}^2\left(2e^{2i\chi}-1\right)\,.
\end{equation}
As a consequence the relation between the cumulants in the SN-case,
$C_n^{\rm SN}$,  and
the normal case, $C_n^{\rm NN}$, is
\begin{equation}
  C_n^{\rm SN}=2^{n-1} C_n^{\rm NN}\,.
  \label{eq:sn-cumulants}
\end{equation}
We observe that we can read off the effective charge from the ratio
$C_n^{\rm SN}/C_n^{\rm NN}$ $=$ $(q_{eff}/e)^{n-1}$ and, indeed, find
$q_{eff}=2e$.  We should emphasize, however, that this is a special
property of the \emph{diffusive connector}. Our prove of
Eq.~(\ref{eq:sn-cumulants}) is valid as long as $eV\ll E_{Th}$, and it
is not clear, wether Eq.~(\ref{eq:sn-cumulants}) is true also for $eV\gg
E_{Th}$

\begin{figure}[tbp]
    \centerline{\psfig{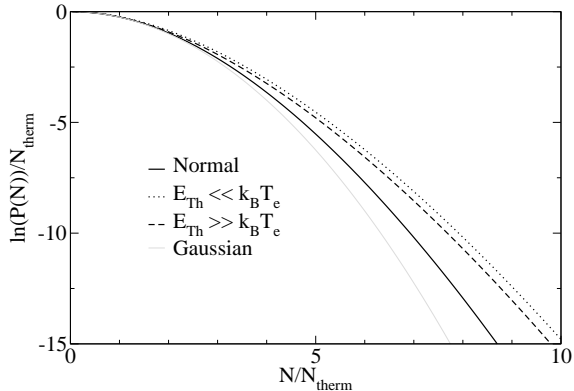}}
    \caption{Equilibrium distribution of the current fluctuations in a
      diffusive SN-contact. We observe a) a strong deviation of all
      distributions from a Gaussian b) enhanced fluctuations in the
      superconducting case, and c) differences between proximity effect
      and coherent transport. Note, that the second cumulant (i.e. the
      thermal noise) is the same for all displayed curves.}
    \label{fig:distr}
\end{figure}

The counting statistics in equilibrium for arbitrary temperature $T_e$
was studied in Ref.~\cite{belzig:01-diff}.  By a numerical solution of
Eq.~(\ref{eq:usadel}), it is possible to evaluate the integral over
$\chi$ in the inversion of Eq.~(\ref{eq:cgf-def}) in the saddle point
approximation, i.e. we take $\chi$ as complex and expand the exponent
around the complex saddle point $\chi=ix_0$.  The integral yields then
$P(N) \approx \exp(-S(ix_0)-x_0 N)$, which we plot implicitly as a
function of $N(x_0)=\partial S(ix_0)/\partial x_0$. Results of this
calculation are displayed in Fig.~\ref{fig:distr}.  The charge number
N is normalized by $N_{\rm therm}=g_dt_0k_BT_e/e^2$. Note that the
conductance (and due to the fluctuation-dissipation theorem the noise)
is the same in all cases. The solid line shows the distribution in the
normal state, which does not depend on the Thouless energy. In our
units, this curve is consequently independent on temperature. In the
superconducting state the Thouless energy does matter, and the
distribution depends on the ratio $E_{\rm Th}/k_BT_e$.  We observe
that large fluctuations of the current in the superconducting case are
enhanced in comparison to the normal case, and in both cases are
enhanced in comparison to Gaussian noise $\sim \exp (-N^2/4N_{\rm
  them})$. The differences between the normal and the superconducting
state occur in the regime of non-Gaussian fluctuations.

\subsection{Supercurrent}
\label{sec:super}

The CGF of a quantum contact, i.~e. Eq.~(\ref{eq:cgf-general}), can be used
to find the counting statistics for a supercurrent between two
superconductors at a fixed phase difference $\phi$. This has been done
in Ref.~\cite{belzig:01-super}. The result can be represented in a form
similar to the CGF of the SN-contact (\ref{eq:cgf-sncontact})
\begin{equation}
  \label{eq:cgf-super}
   S(\chi,\phi)=-\frac{t_0}{2\pi}\sum_n\int dE
  \ln\left[1+\sum\limits_{q=-2}^{2}
  \frac{A_{n,q}^S(E,\phi)}{Q_n(E,\phi)}\left(e^{iq\chi}-1\right)\right]\,. 
\end{equation}
Explicit expressions for the coefficients are given in Appendix
\ref{sec:coefficients-ss}. To find the
statistics of the charge transfer, we will treat two separate cases.

{\em Gapped superconductors.} If the two leads are gapped like BCS
superconductors, the spectral functions are given by
Eq.~(\ref{eq:green-super}). Here we account for a finite lifetime
$\delta$ of the Andreev bound states, {\em e.~g.}  due to phonon
scattering. The supercurrent in a one-channel contact of transparency
$T_1$ is solely carried by Andreev bound states with energies
$\pm\Delta(1-T_1\sin^2\phi/2)^{1/2} \equiv \pm E_B(\phi)$.  The
importance of these bound states can be seen from the coefficient
$Q(E,\phi)$ (see Eq.~(\ref{eq:coeffq})), which may become zero and
will thus produce singularities in the CGF.\footnote{It is interesting
  to note that in the limit $\delta\to 0$ the CGF has poles for
  energies $E_B^2(\chi) = \Delta^2 \left(1-T \sin^2\left(
      \frac{\phi\pm\chi}{2} \right)\right)$.  The counting field
  therefore couples directly to the phase sensitivity of the Andreev
  bound states.} The broadening $\delta$ shifts the singularities of
$Q(E,\phi)$ into the complex plane and allows an expansion of the
coefficients $A_{1,q}^S$ close to the bound state energy.  Performing
the energy integration the CGF results in
\begin{equation}
  \label{eq:bs-action1}
  S(\chi,\phi)=-2t_0\delta\sqrt{1-\chi^2\frac{I_1^2(\phi)}{4\delta^2}
    -i\chi \frac{I_1(\phi)}{\delta}
    \tanh\left(\frac{E_B(\phi)}{2k_BT_e}\right)}\,,
\end{equation}
where $I_1(\phi)=\Delta^2 T_1 \sin(\phi)/2E_B(\phi)$ is the
supercurrent carried by one bound state. In deriving
(\ref{eq:bs-action1}) we have also assumed that $\chi\ll 1$. This
corresponds to a restriction to ``long trains'' of electrons
transfered, and the discreteness of the electron transfer plays no
role here.  Fast switching events become less probable at low
temperatures and are neglected here.  In the saddle point
approximation and for $\gamma(\phi)\equiv1/\cosh(E_B(\phi)/2T_e)\ll 1$
the FCS can be found. We express the transfered charge in terms of the
current normalized to the zero temperature supercurrent:
$j(\phi)=eN/t_0I_1(\phi)$. We find for the current distribution in the
saddle point approximation
\begin{equation}
  \label{eq:super-distr}
  P(j,\phi)\sim\frac{1}{\gamma}
  \exp\left[2\delta t_0\left(\gamma(\phi)\sqrt{1-j^2(\phi)} 
      -j(\phi)\sqrt{1-\gamma^2(\phi)}\right)\right])
  \,,
\end{equation}
for $|j(\phi)|\le 1$ and zero otherwise.  At zero temperature
Eq.~(\ref{eq:super-distr}) approaches $P(j,\phi)\to\delta(j-1)$.  Thus
the charge transfer is noiseless. At finite temperature, on the other
hand, the distribution (\ref{eq:super-distr}) confirms the picture of
switching between Andreev states which carry currents in opposite
directions, suggested in Ref.~\cite{averin:96}. The previous result is
valid under the following conditions. In the energy integration it was
assumed that the bound states are well defined. For small transmission
the distance of the bound state to the gap edge is $\approx T_1\Delta$.
Thus, to have well-defined bound states we have to require $\delta <
T_1\Delta$. Similarly, for a highly transmissive contact and a small 
phase difference we require $\phi\sim eN/t_0I_{\rm c} > \delta/\Delta$. The
statistics beyond these limits is similar to what is discussed in the
following.

{\em Tunnel junction/gapless superconductors.} Let us now consider the
supercurrent statistics between two weak superconductors, where the
Green's functions can be expanded in $\Delta$ for all energies. One can
see that this is equivalent to the tunnel result
(\ref{eq:cgf-tunnel}). It has the form
\begin{equation}
  \label{eq:cgf-super-tunnel}
  S(\chi,\phi) = -N_{+}(\phi)(e^{i\chi}-1)-N_{-}(\phi)(e^{-i\chi}-1)\,.
\end{equation}
where
\begin{equation}
  \label{eq:coeff-super-tunnel}
  N_{\pm}(\phi)=\frac{t_0}{2}\left(P_{\rm s}(\phi) \pm I_{\rm s}(\phi)\right)\,.
\end{equation}
This form of the CGF shows that the FCS is expressed in terms of
supercurrent $I_{\rm s}(\phi)$ and noise $P_{\rm s}(\phi)$
only. Supercurrent and current noise are
\begin{eqnarray}
  \label{eq:super-tunnel-current}
  I_{\rm s}(\phi) = -\frac{G_T}{4}\textrm{Re}\int dE 
  \textrm{Tr}\left\{\hat\sigma_3\left[\hat R_1(E,\phi),\hat
      R_2(E)\right]\right\}
  \tanh\left(\frac{E}{2k_BT_e}\right)\,,
\\
  \label{eq:super-tunnel-noise}
  P_{\rm s}(\phi) =-\frac{G_T}{4}\textrm{Re}\int dE 
  \textrm{Tr}\left\{\hat\sigma_3\hat A_1(E,\phi)\hat\sigma_3\hat
    R_2(E)\right\}
  \frac{1}{\cosh^{2}\left(\frac{E}{2k_BT_e}\right)}\,.
\end{eqnarray}
Here $G_T$ is the normal-state conductance of the contact.  Eq.
(\ref{eq:super-tunnel-noise}) shows that $P_{\rm s}$ vanishes at zero
temperature, whereas $I_{\rm s}$ vanishes at $T_{\rm c}$.  Therefore,
there is some crossover temperature below which $P_{\rm s}<|I_{\rm
  s}|$.  In this limit one of the coefficients $N_{\pm}$ becomes
negative and the interpretation, that the CGF
(\ref{eq:cgf-super-tunnel}) corresponds to a generalized Poisson
distribution, makes no sense anymore.  In fact, the CGF leads to
'negative probabilities' and does not correspond to any probability
distribution, The origin of this failure is the broken U(1)-symmetry
the superconducting state. Nevertheless the FCS can be used to predict
the outcome of any charge transfer measurement, as is discussed in
detail in Refs.~\cite{belzig:01-super,kindermann:01,shelankov}.

\section{Multi-Terminal Structures}
\label{cha:multi}

\begin{figure}[bth]
  \centerline{\psfig{figure=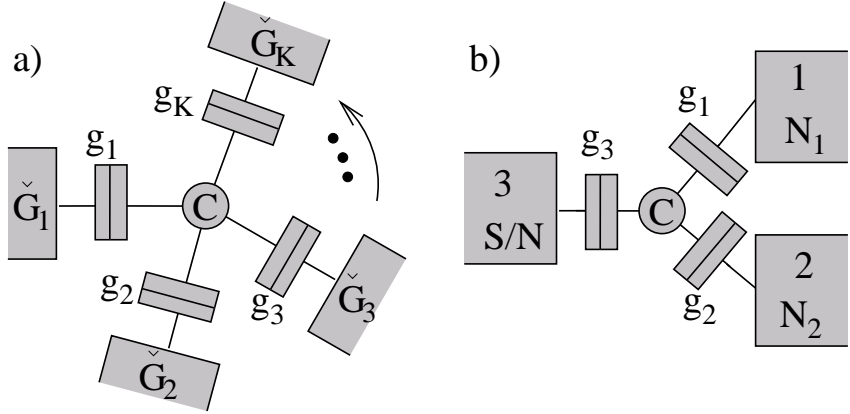,width=0.9\textwidth}}
  \caption{a) An example for a multi-terminal structure. $K$ terminals
    are connected to a central node by tunnel junctions b)
    Three-terminal structure. Two different voltage configurations are
    considered. In the first case, all terminals are normal metals. A
    bias voltage is applied between terminal 1 and terminal 2, whereas
    terminal 3 is operated as a voltage probe (no mean current). In the
    second case we consider a beam splitter configuration. A
    supercurrent or normal current in terminal 3 is divided in (or
    merged from) two normal currents from terminals 1 and 2. Here, we
    assume that the same potential is applied to terminals 1 and 2.}
  \label{fig:multi-terminal}
\end{figure}

Many mesoscopic transport experiments are performed in multi-terminal
configurations. An example is shown in Fig.~\ref{fig:multi-terminal}.
Due to the quantum nature of the charge carriers, interesting non-local
effects can appear, such as sensitivity of measured voltage differences
to changes in the setup outside the current path.  Obviously, the same
is true for current fluctuations and for the full counting statistics.
These are sensitive to the quantum correlations between the charge
carriers, which can have a nonlocal character. In terms of counting
statistics this means, for example, that the joint probability to count
$N_1$ particles in terminal 1 and $N_2$ particles in terminal 2 can not
be factorized into separate probabilities for the two events.

\subsection{General Result for Multi-Tunnel Geometry}
\label{sec:multi-theory}

The generalization of the method, introduced in Section \ref{cha:theory}
for two terminals, to many terminals is straightforward
\cite{yuli:01-multi}. The counting field $\chi$ is replaced by a vector
$\vec\chi$, with dimension equal to the number of terminals. For brevity
we collect the charges passing each terminal into a vector $\vec N$. The
current in each terminal is coupled to the respective component of the
counting field by an expression like (\ref{eq:hpert}).  Following the
procedure outlined in Sections \ref{sec:fcs-theory} and
\ref{sec:circuit} the result is,  that the Green's function of each
terminal acquires its own counting field $\chi_n$. The rules, that
determine the transport properties, remain essentially unchanged.

The procedure outlined above is best illustrated by an example. We
consider a node connected to $K$ terminals via tunnel junctions with
conductances $g_k$. This setup is shown in
Fig.~\ref{fig:multi-terminal}a. Each terminal is described by a Green's
function $\check G_k(\chi_k)$ ($k=1..K$), which is related to the
terminal's usual Green's function by a counting rotation
(\ref{eq:countrot}). We do not need to specify yet, whether the terminals
are superconducting or normal. The goal is to find (for arbitrary
applied voltages and temperatures) the joint probability
$P_{t_0}(N_1,N_2,...,N_K)$, that $N_1$ particles enter through terminal
1, $N_2$ particles through terminal 2, ..., and $N_K$ particles through
terminal $K$.  Correspondingly we define the cumulant generating
function
\begin{equation}
  \label{eq:multi-cgf}
  e^{-S(\vec \chi)}
  =\sum_{N_1,N_2,\cdots,N_K} P_{t_0}(\vec N)
  e^{i \vec N\vec \chi}\,.
\end{equation}
The central node is described by a Green's function $\check
G_c(\vec\chi)$, which has to be determined from the circuit rules. The
matrix currents through terminal $k$ is given by $\check
I_k=-(g_k/2)[\check G_k(\chi_k) , \check G_c(\vec\chi)]$ and current
conservation on the node can be written as
\begin{equation}
  \label{eq:multi-currentconserve}
  0=\sum_{k=1}^{K} \check I_k=
  \frac12 \left[\sum_{k=1}^{K} g_k\check G_k(\chi_k), 
    \check G_c(\vec\chi)\right]\,.
\end{equation}
This equation (together with the normalization condition $\check
G_c^2(\vec\chi)=\check 1$) is solved by
\begin{equation}
  \label{eq:multi-solution}
  \check G_c(\vec\chi)=\frac{\sum_{k=1}^{K} g_k\check G_k(\chi_k)}{
    \sqrt{
      \sum_{k,l=1}^{K} g_kg_l
      \left\{\check G_k(\chi_k),\check G_l(\chi_l)\right\}}}\,.
\end{equation}
The CGF is found from integrating the relations $\partial
S(\vec\chi)/\partial\chi_k=(-it_0/e)$ $ I_k(\vec\chi)$, where
$I_k(\vec\chi)=(1/4e)\int dE $Tr$\check\tau_K\check I_k(\vec\chi)$.  The
CGF is then determined up to an additive constant, which is fixed by the
normalization $S(\vec 0)=0$ and neglected in the following. The result is
\cite{boerlin-prl}
\begin{equation}
  \label{eq:cgf-multi-tunnel}
  S(\vec\chi)=-\frac{t_0}{4e^2}\int dE \;\textrm{Tr}
  \sqrt{\sum\nolimits_{k,l=1}^{K} g_kg_l\left\{\check G_k(\chi_k) , \check
        G_l(\chi_l)\right\}}\,.
\end{equation}
This provides the
counting statistics for an arbitrary multi-terminal structure of the
type shown in Fig.~\ref{fig:multi-terminal}. Below we discuss several
examples.

\subsection{Normal Metal Multi-Terminal Structures}
\label{sec:multi-normal}

In the case that all terminals are normal metals, we can evaluate
Eq.~(\ref{eq:cgf-multi-tunnel}) further. All terms under the square root
are proportional to the unit matrix and the trace can be taken easily
with the result \cite{boerlin-diplom}
\begin{eqnarray}
  \label{eq:cgf-multi-normal}
    S(\vec\chi) & = & -\frac{t_0g_\Sigma}{2e^2}\int dE\times
  \\\nonumber&& 
  \sqrt{1+ \sum\nolimits_{k,l=1}^{K} t_{kl}
    \left[f_k(1-f_l)e^{i(\chi_k-\chi_l)}+
      f_l(1-f_k)e^{i(\chi_l-\chi_k)}\right]}\,,
\end{eqnarray}
where $g_\Sigma=\sum_{k=1}^K g_k$ and $t_{kl}=2g_kg_l/g_\Sigma^2$.  The
argument of the square root is the sum over all tunneling events from
between the terminals. For two terminals we recover the result
(\ref{eq:cgf-dbl-normal}).

Let us now consider three normal-metal terminals, one of which is
operated as voltage probe, i.e. no average current enters the terminal.
This layout is depicted in Fig.~\ref{fig:multi-terminal}. We assume for
the applied potentials that $V_1<V_3<V_2$. Terminal 3 is operated as
a voltage probe and it follows, that $V_3=(g_1V_1+g_2V_2)/(g_1+g_2)$. At
zero temperature the energy integration yields
\begin{eqnarray}
  \label{eq:cgf-trpl-normal}
  S(\vec\chi) & = & -\frac{t_0g_\Sigma V}{2e}
  \left[\frac{g_2}{g_1+g_2}
    \sqrt{1+t_{13} (e^{i(\chi_3-\chi_1)}-1) 
      +t_{12}(e^{i(\chi_2-\chi_1)}-1)}
  \right.\nonumber\\
  &&\left.
    +\frac{g_1}{g_1+g_2}
    \sqrt{1+t_{23} (e^{i(\chi_2-\chi_3)}-1) 
      +t_{12}(e^{i(\chi_2-\chi_1)}-1)}
  \right]\,. 
\end{eqnarray}
The CGF
separates into two terms.  The first term corresponds to the energy
window $eV_3>E>eV_1$, in which transport is only possible between
terminals 1 and 3 or between 1 and 2.  The second term results from the
energy window $eV_2>E>eV_3$, in which no electrons can enter into
terminal 2.

We now consider a different configuration: an instreaming current is
divided into two outgoing currents. This corresponds to the voltage
configuration $V=V_1=V_2>V_3=0$. The corresponding CGF is
\begin{equation}
  \label{eq:cgf-splitter-normal}
  S(\vec\chi)=-\frac{t_0g_\Sigma V}{2e}
  \sqrt{1+t_{13} (e^{i(\chi_1-\chi_3)}-1) 
      +t_{23}(e^{i(\chi_2-\chi_3)}-1)}\,.
\end{equation}
In the limit $g_2,g_1\gg g_3$ or vice versa the CGF takes the form
\begin{equation}
  \label{eq:cgf-splitter-normal1}
  S(\vec\chi)=-\frac{t_0g_\Sigma V}{4e}
  \left(t_{13} (e^{i(\chi_1-\chi_3)}-1) 
    +t_{23}(e^{i(\chi_2-\chi_3)}-1)\right)\,,
\end{equation}
and the corresponding counting statistics is
\begin{equation}
  \label{eq:fcs-splitter-normal}
  P(N_1,N_2)=e^{-(\bar N_1+\bar N_2)}
  \frac{{\bar N_1}^{N_1}}{N_1!}
  \frac{{\bar N_2}^{N_2}}{N_2!}\,.
\end{equation}
Here $\bar N_i=(t_0V/e) g_ig_3/4g_\Sigma^2$. The statistics is a product
of two Poisson distributions, i.~e. the two transport processes are
uncorrelated.

\subsection{Superconducting Multi-Terminal Structures}
\label{sec:multi-super}

Let us now consider the beam splitter configuration, if the incoming current
originates from a superconductor. Here, we have to use the
Nambu$\times$Keldysh matrix structure. The layout is as shown in
Fig.~\ref{fig:multi-terminal}. We choose terminal 3 as superconducting
terminal with $V_3=0$ and the potential in the two normal terminal is assumed
to be the same, $V_1=V_2=V$. We consider the limit $T\ll eV \ll
E_{Th},\Delta$.  Transport occurs then only via Andreev reflection, since no
quasiparticles in the superconductor are present.  The Green's functions for
the terminals can be found in Appendix \ref{sec:bulk-solutions}. We note, that
the pair breaking effect due to a magnetic field in a chaotic dot with the
same terminal configuration was studied by Samuelsson and B\"uttiker
\cite{samuelsson1}.

We now evaluate Eq.~(\ref{eq:cgf-multi-tunnel}) for the three-terminal
setup. The CGF depends only on the differences $\chi_1-\chi_3$ and $
\chi_2-\chi_3$, which is a consequence of charge conservation and allows
to drop the explicit dependence on $\chi_3$ below.  Introducing $p_i=2
g_3 g_i /(g_3^2+(g_1+g_2)^2)$ we find
\begin{eqnarray}
 \label{eq:cgf-trpl-super}
 \lefteqn{S(\chi_1,\chi_2)=
   -\frac{Vt_0\sqrt{g_3^2+(g_1+g_2)^2}}{\sqrt2e}\times}\\\nonumber
 &&
 \sqrt{1+\sqrt{1+
 p_1^2 (e^{i2\chi_1}-1)+
 p_2^2 (e^{i2\chi_2}-1)+
 2p_1p_2 (e^{i(\chi_1+\chi_2)}-1)}}\,.
\end{eqnarray}
The inner argument contains counting factors for the different possible
processes. A term $\exp(i(\chi_k+\chi_l)-1)$ corresponds to an event in
which two charges leave the superconducting terminal and one charge is
counted in terminal $k$ and one charge in terminal $l$. The prefactors
are related to the corresponding probabilities. For instance, $p_1$ is
proportional to the probability of a coherent tunneling event of an
electron from the superconductor into terminal 1. A coherent
pair-tunneling process is therefore weighted with $p_1^2$. This is
accompanied by counting factors which describe either the tunneling of
two electrons into terminal 1(2) (counting factor
$\exp(i(2\chi_{1(2)}))-1)$) or tunneling into different terminals
(counting factor $\exp(i(\chi_1+\chi_2)-1)$).  The nested square-root
functions show that these different processes are non-separable.

It is interesting to consider the limiting case if $g_3/(g_1+g_2)$ is not
close to 1. Then, $p_{1,2}\ll 1$ and we can expand
Eq.~(\ref{eq:cgf-trpl-super}) in $p_{1,2}$. The CGF can be written as
\begin{eqnarray}
 \label{eq:cgf-trpl-super1}
 \lefteqn{S(\chi_1,\chi_2) = 
   -\frac{t_0V}{e}\frac{g^2_3}{(g^2_3+(g_1+g_2)^2)^{3/2}}\times}
 \\\nonumber&&
 \left[g_1^2 (e^{i2\chi_1}-1)+ g_2^2 (e^{i2\chi_2}-1)
   +2 g_1g_2(e^{i(\chi_1+\chi_2)}-1) \right] \; .
\end{eqnarray}
The CGF is composed of three different terms, corresponding to a charge
transfer event of $2e$ either into terminal 1 or terminal 2 (the first two
terms in the bracket) or separate charge transfer events into terminals 1 and
2. The same form of the CGF appears if the proximity effect is destroyed by
other means, e.~g. a magnetic field \cite{samuelsson1}. According to the
general principles of statistics, sums of CGFs of independent statistical
processes are additive.  Therefore, the CGF (\ref{eq:cgf-trpl-super1}) is a
sum of CGFs of independent Poisson processes.  The total probability
distribution $P(N_1,N_2)$ corresponding to Eq.~(\ref{eq:cgf-trpl-super1}) can
be found. It vanishes for odd values of $(N_1+N_2)$ and for even values it is
given by
\begin{equation}
 \label{eq:fcs-trpl-super-poisson}
 P(N_1,N_2) = 
  \frac{e^{-\frac{\bar N}{2}}
    \left(\frac{\bar N}{2}\right)^{\frac{N_1+N_2}{2}}}{
    \left(\frac{N_1+N_2}{2}\right)!}
  {{N_1+N_2}\choose{N_1}}
  T_1^{N_1}T_2^{N_2}\,.
\end{equation}
Here we have defined the average number of transfered electrons $\bar N
= (t_0 V/e) (g_1+g_2)^2 g_3^2/((g_1+g_2)^2+g_3^2)^{3/2}$ and the
probabilities $T_{1(2)}=g_{1(2)}/$ $(g_1+g_2)$ that one electron leaves the
island into terminal $1(2)$. If one would not distinguish electrons in
terminals 1 and 2, the charge counting distribution can be obtained from
Eq.~(\ref{eq:cgf-trpl-super1}) by setting $\chi_1=\chi_2=\chi$ and
performing the integration. This leads to $P^S_{tot}(N)=\exp(-\bar N/2)
(\bar N/2)^{N/2}/(N/2)!$, which corresponds to a Poisson distribution of
an uncorrelated transfer of electron pairs. The full distribution
Eq.~(\ref{eq:fcs-trpl-super-poisson}) is given by $P^S_{{tot}}(N_1+N_2)$,
multiplied with a \textit{partitioning factor}, which corresponds to the
number of ways to distribute $N_1+N_2$ identical electrons
among the terminals 1 and 2, with respective probabilities $T_1$ and
$T_2$. Note, that $T_1+T_2=1$, since the electrons have no other
possibility to leave the island.

\section{Concluding Remarks}
\label{cha:conc}

Full Counting Statistics is a new fundamental concept in mesoscopic
electron transport. The knowledge of the full probability distribution
of transfered charges completes the information on the transport
mechanisms. In fact, the FCS represents \emph{all} information, which
can be gained from charge counting in a transport process - a clear
progress in our understanding of quantum transport.

In this article we have reviewed the state of the field with particular
emphasis on superconductor-normal metal heterostructures. We have
introduced a theoretical method, which combines full counting statistics
with the powerful Keldysh Green's function technique. This method allows
to obtain the FCS for a large variety of mesoscopic systems. For a
two-terminal structure a general relation can be derived which contains
the FCS of all kinds of constrictions between normal metals and
superconductors. Our method is readily applicable to multi-terminal
structures and we have discussed one example. An important
advantage of the method is that it allows a direct numerical
implementation, which means that we are able to find the FCS of
arbitrary mesoscopic SN-structures.

I would like to thank the Alexander von Humboldt-Stiftung, the
Dutch FOM, the Swiss SNF, and the NCCR Nanoscience for financial support
in different stages of this work. Special thanks go to Yuli V. Nazarov,
who introduced many of the concepts discussed here. Valuable insights
emerged from discussions with D. Bagrets, J.  B\"orlin, C. Bruder, M.
B\"uttiker, M. Kindermann, P.  Samuelsson, and F. Taddei.

\appendix
\section{Field Theoretical Methods}
\label{sec:keldysh}

We summarize some methods and definitions of quantum field theory. In
the first part we review briefly the standard Keldysh-Green's function
technique.  We follow essentially the review \cite{rammer}. In the
second part we explicitly perform the linked cluster expansion for the
cumulant generating function, which establishes the relation between our
definition of FCS and the Green's function method.

\subsection{Keldysh Green's Functions}
\label{sec:keldysh-green}

\begin{figure}[htbp]
    \centerline{\psfig{figure=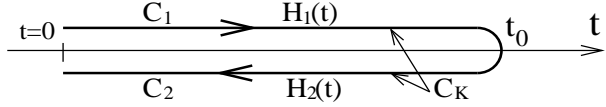,width=8cm}}
    \caption[]{The Keldysh time ordering contour $C_K$}
    \label{fig:timepath}
\end{figure}

One commonly used method to study nonequilibrium phenomena is the
so-called Keldysh technique. Quite generally time-dependent problems are
cast into the form of calculating expectation values of some operator
$A$ of the form $\langle A(t) \rangle$, where $A(t)$ is the
time-dependent operator in the Heisenberg picture with respect to some
Hamiltonian: $A(t)=U^\dagger(t,t^\prime)A(t^\prime)U(t,t^\prime)$. In
this expression both the time-ordered evolution operator $U(t,t^\prime)$
$=$ ${\cal T}$ $\exp(-i\int_{t^\prime}^t d\tau H(\tau))$ and the
anti-time-ordered evolution operator $U^\dagger(t,t')$ $=$ $ \tilde{\cal
  T} \exp(-i\int^{t^\prime}_t d\tau H(\tau))$ appear. A diagrammatic
theory requires to account for the various time orderings, which is
rather complicated.  A considerable simplification arises from Keldysh's
trick. We introduce two time coordinates $(t_1,t_2)$, which live on the
upper and lower part of the contour $(C_1,C_2)$, and an ordering
prescription along the \textrm{closed time path} $C_K$, depicted in
Fig.~\ref{fig:timepath}.

In the context of full counting statistics we introduce \emph{different}
Hamiltonians $H_{1(2)}$ for the two parts of the contour $C_{1(2)}$. The
actual different parts are given by Eq.~(\ref{eq:hpert}). The rest of the
Hamiltonian $H_{\rm\small sys}$ coincides on both parts, just
as in the usual formulation \cite{rammer}.

We define a \textit{contour-ordered} Green's function (${\cal T}_{K}$
denotes ordering along the Keldysh-contour)
\begin{equation}
  \hat G_{C_K}(t,t^\prime)=
  -i\langle {\cal T}_{K}\Psi(t)\Psi^\dagger(t^\prime)\rangle\,.
\end{equation}
Here the field operators can have multiple components, such as spin or
Nambu for example.  This Green's function can be mapped onto a matrix
space, the so-called Keldysh space, by considering the time coordinates
on the upper and lower part of the Keldysh contour as formally
independent variables
\begin{eqnarray}
  \check G(t,t^\prime)&\equiv&
  \left( 
    \begin{array}[c]{rr}
      \hat G_{11}(t,t^\prime) & \hat G_{12}(t,t^\prime)\\
      -\hat G_{21}(t,t^\prime) & -\hat G_{22}(t,t^\prime)
    \end{array}\right)\,,
  \label{eq:keldysh-green}
\end{eqnarray}
which reads in terms of the field operators in the Heisenberg picture
\begin{eqnarray}
  -i\left(
    \begin{array}[c]{rr}
      \langle {\cal T}\Psi_{H_1}^{\phantom{\dagger}}(t)\Psi_{H_1}^\dagger(t')\rangle &
      \langle \Psi_{H_1}^\dagger(t')\Psi_{H_2}^{\phantom{\dagger}}(t)\rangle\\
      -\langle \Psi_{H_2}^{\phantom{\dagger}}(t)\Psi_{H_1}^\dagger(t')\rangle &
      \langle \tilde{\cal T}\Psi_{H_2}^{\phantom{\dagger}}(t)\Psi_{H_2}^\dagger(t')\rangle
    \end{array}\right)\,.
\end{eqnarray}
The current is obtained from the Green's functions (with spatial
coordinates reinserted) as
\begin{eqnarray}
  \hat{\mathbf{j}}({\mathbf x},t)&=&-\frac{e}{2m}\lim\limits_{
      {\mathbf x}\to {\mathbf x}^\prime}
  \left(\nabla_{{\mathbf x}}-\nabla_{{\mathbf x}^\prime}\right)
  \hat G_{12}({\mathbf x},t;{\mathbf x}^\prime,t)
  \\\nonumber
  &=&
  -\frac{e}{2m}\lim\limits_{
      {\mathbf x}\to {\mathbf x}^\prime}
  \left(\nabla_{{\mathbf x}}-\nabla_{{\mathbf x}^\prime}\right)
  (\hat G_{11}({\mathbf x},t;{\mathbf x}^\prime,t)
  +\hat G_{22}({\mathbf x},t;{\mathbf x}^\prime,t))\,,
\end{eqnarray}
where the second form is the one we have used in the context of counting
statistics. The current has still a matrix structure in the subspace of
the components of the field operators. How the electric current is
obtained, depends on the definition of the subspace.

In the usual Keldysh-technique $H_1=H_2$ and we have the general
property $G_{11}-G_{22}=G_{12}-G_{21}$. One element of the Green's
function (\ref{eq:keldysh-green}) can be eliminated by the transformation
$\underline G={\cal L}\check G {\cal L}^\dagger$, where ${\cal
  L}=(1-i\bar\tau_2)/\sqrt 2$ and $\bar \tau_i$ denote Pauli matrices in
Keldysh space. Then the matrix Green's function takes the form
\begin{equation}
  \label{eq:keldysh-matrix}
  \underline G(t,t^\prime)=\left(
    \begin{array}[c]{cc}
      \hat G^R(t,t^\prime) & \hat G^K(t,t^\prime)\\
      0 & \hat G^A(t,t^\prime)
    \end{array}\right)\,,
\end{equation}
where
\begin{eqnarray}
  \hat G^{R(A)}(t,t^\prime)&=&\mp i\theta(\pm(t-t^\prime))
  \langle[\Psi(t),\Psi^\dagger(t^\prime)]\rangle\,,
  \\
  \hat G^K(t,t^\prime)&=&-i
  \langle\{\Psi(t),\Psi^\dagger(t^\prime)\}\rangle\,.
\end{eqnarray}
The bulk solutions for normal metals and superconductors in
App.~\ref{sec:bulk-solutions} are given in this form.

\subsection{The Relation between Cumulant Generating Function and Green's 
  Function}
\label{sec:cluster}

We establish the connection between the diagrammatic expansions of the
Green's function, defined in (\ref{eq:eom}), and the cumulant generating
function in (\ref{eq:cgf-general}).  We first show how the diagrammatic
expansion of the CGF is obtained. In fact, the same  expansion occurs in
the expression for the thermodynamic potential \cite{agd}. According to
the definition of the CGF (\ref{eq:cgf-general}) we write
\begin{equation}
  \label{eq:cgf-potential}
  e^{-S(\chi)}=\langle T_{C_K} e^{-i\frac{\chi}{2e}\int_{C_K}dtI_c(t)}\rangle\,.
\end{equation}
We consider a term of the order $n$ in the expansion of the
exponent. Such a term has a form
\begin{equation}
  \label{eq:cluster1}
  \frac{1}{n!}\left(-i\frac{\chi}{2e}\right)^n
  \int_{C_K}\cdots\int_{C_K}dt_1\cdots dt_n
  \langle T_{C_K}I_c(t_1) I_c(t_2)\cdots I_c(t_n)\rangle\,.
\end{equation}
We abbreviate for the sum of connected diagrams of order n
\begin{equation}
  \label{eq:cluster-connected}
  Q_n=  \frac{1}{n!}
  \int_{C_K}\cdots\int_{C_K}dt_1\cdots dt_n
  \langle T_{C_K}I_c(t_1) I_c(t_2)\cdots I_c(t_n)\rangle_{con}\,.
\end{equation}
To count the possible diagrams, which contain $p_1$ connected
diagrams of order $m_{1}$, $p_2$ of order $m_2$, and so on, where
$p_1m_1+p_2m_2+\cdots+p_km_k=n$, we note that their number is equivalent
to the number of possibilities to assign $n$ operators to $p_1$ cells containing
$m_1$ places, $p_2$ cells containing $m_2$ places, ... . This number is
given by
\begin{equation}
  \label{eq:cgf-number}
  \frac{n!}{p_1! (m_1!)^{p_1} p_2! (m_2!)^{p_2} \cdots p_k! (m_k!)^{p_k}}
\end{equation}
and it follows that the CGF can be written as
\begin{eqnarray}
  \label{eq:cgf-summation}
  e^{-S(\chi)} & = &
  \sum_{p_1,p_2,\cdots} 
  \frac{1}{p_1!}\left[-i\frac{\chi}{2e}Q_1\right]^{p_1}\frac{1}{p_2!}
  \left[\left(-i\frac{\chi}{2e}\right)^2 Q_2\right]^{p_2} \cdots\\
  &=& \exp\left[-i\frac{\chi}{2e}Q_1+\left(-i\frac{\chi}{2e}\right)^2 Q_2+\cdots\right]\,.
\end{eqnarray}
Thus, we find that the CGF $S(\chi)$ is directly given by the sum
\begin{equation}
  \label{eq:cgf-summation2}
  -S(\chi) = \sum_{n=1}^\infty \left(i\frac{\chi}{2}\right)^n Q_n 
\end{equation}
over the connected diagrams only.

The Green's function introduced in Eq.~(\ref{eq:eom}) has a perturbation
expansion
\begin{eqnarray}
  \label{eq:green-perturbation}
  G({\mathbf x},t;{\mathbf x}',t',\chi) & = & 
  -i\sum_{n=0}^{\infty}\frac{1}{n!} 
    \left(-i\frac{\chi}{2e}\right)^n
    \int_{C_K}\cdots\int_{C_K}dt_1\cdots dt_n\times
    \\\nonumber &&
    \langle T_{C_K}\Psi({\mathbf x},t)\Psi^\dagger({\mathbf x}',t')
    I_c(t_1) I_c(t_2)\cdots I_c(t_n)\rangle_{con}\,,
\end{eqnarray}
recalling that all disconnected diagrams are canceled from this
expression.  Now we calculate the current from this Green's function
with the same current operator used in $I_c$. We find
\begin{eqnarray}
  \label{eq:eom-chi-current}
  I(\chi,t)&=& -\sum_{n=1}^{\infty}\frac{1}{(n-1)!} 
    \left(-i\frac{\chi}{2e}\right)^{n-1}
    \int_{C_K}\cdots\int_{C_K}dt_1\cdots dt_{n-1}\times
    \\\nonumber &&
    \langle T_{C_K} I_c(t_1) I_c(t_2)\cdots I_c(t_{n-1})I_c(t)
    \rangle_{con}\,.
\end{eqnarray}
In a static situation (as we consider) this current does not depend on
$t$ and we integrate along the Keldysh contour between 0 and
$t_0$. Using Eq.~(\ref{eq:cluster-connected}) it follows that
\begin{eqnarray}
  \label{eq:eq:eom-chi-current2}
  2t_0I(\chi) & = & - \sum_{n=1}^{\infty} n 
    \left(-i\frac{\chi}{2e}\right)^{n-1} Q_n
\end{eqnarray}
By comparing the right-hand sides of (\ref{eq:eq:eom-chi-current2}) and
(\ref{eq:cgf-summation2}) we finally obtain the relation between the
$\chi$-dependent current and the CGF:
\begin{equation}
  \label{eq:cgf-current-relation}
  I(\chi) = i\frac{e}{t_0} \frac{\partial S(\chi)}{\partial \chi}\,.
\end{equation}
The constant contribution to $S(\chi)$ follows from the normalization
$S(0)=0$.

\section{Quasiclassical Approximation}

In practice an exact calculation of Green's functions is impossible in
virtually all mesoscopic transport problems. An important
simplification is the quasiclassical approximation \cite{eilenberger},
which makes use of the smallness of the most energy scales involved in
transport with respect to the Fermi energy $E_F$.  We briefly
summarize the derivation of the basic equations \cite{rammer}. Here we
concentrate on the derivation in the context of superconductivity. The
inclusion of spin-dependent phenomena is straightforward.

\subsection{Eilenberger Equation}
\label{sec:eilenberger}

The starting point is the equation of motion for the real-time
single-particle Green's function. We consider here the static case, in
which the equations can be considered in the energy representation. The
equation of motion reads
\begin{equation}
  \label{eq:gorkov}
  \left[E\hat\sigma_3-\frac{{\mathbf p}^2}{2m}+E_F-\check\Sigma\right] 
  \check G({\mathbf x}, {\mathbf x'};E) = 
  \delta({\mathbf x}-{\mathbf x}')\,.
\end{equation}
Here the $\check{ }$ denotes matrices in the combined
Nambu$\times$Keldysh-space and we use $\hat\sigma_i(\bar\tau_i)$ to
denote Pauli matrices in Nambu(Keldysh)-space. $E_F=p_F^2/2m$ is the
Fermi energy and the selfenergy $\hat\Sigma$ includes scattering
processes. The complicated dependence on two spatial coordinates in
Eq.~(\ref{eq:gorkov}) can be eliminated by the following procedure. We
introduce the Wigner transform
\begin{equation}
  \check
  G({\mathbf{r}},{\mathbf{p}};E)=\int d^3s \exp\left(i{\mathbf{ps}}\right)
  \check{G}({\mathbf r}+{\mathbf s}/2,{\mathbf r}-{\mathbf s}/2;E)\,,
\label{eq:quasi-classical-gfunc}
\end{equation}
for which the equation of motion reads
\begin{equation}
  \label{eq:wigner}
  \left[E\hat\sigma_3 -i{\mathbf{v}}
    \nabla_{{\mathbf r}} 
    -\check\Sigma({\mathbf r},{\mathbf{p}})\right] 
  \check G({\mathbf r},{\mathbf p},E)
  =\check 1\,.
\end{equation}
In this equation, we can neglect the dependence on the absolute value of
the momentum in the expression in the brackets, since $\check G({\mathbf
  r},{\mathbf p},E)$ is strongly peaked at ${\mathbf p}={\mathbf p_F}$.
We now subtract the inverse equation and integrate the resulting
equation over $\xi=p^2/2m$.  We are lead to the definition of the
quasiclassical Green's function
\begin{equation}
  \label{eq:quasicl-greensfunction}
  \check g({\mathbf{r}},{\mathbf{p_F}},E) = \frac{i}{\pi} 
  \int \frac{dp p}{m} \check G({\mathbf r},{\mathbf p};E)\,.
\end{equation}
The quasiclassical Green's function obeys the Eilenberger
equation \cite{eilenberger}
\begin{equation}
  \label{eq:eilenberger}
  \frac{1}{e^2N_0}\nabla \check{{\mathbf{j}}}({\mathbf{r}},{\mathbf{v_F}},E)
  =
  \left[-iE\hat\sigma_3+i\check\sigma({\mathbf r},{\mathbf v_F})\,,\,
    \check g({\mathbf r},{\mathbf v_F};E)\right]\,.
\end{equation}
The current density is obtained from the generalized matrix-current
\begin{eqnarray}
  \label{eq:quasi-current-clean}
  \check{{\mathbf{j}}}({\mathbf{r}},{\mathbf{v_F}},E) &=&
  e^2N_0{\mathbf{v_F}}\check g({\mathbf{r}},{\mathbf{v_F}},E)\\
  {\mathbf j}({\mathbf r})&=& \frac{1}{4e} 
  \int dE \textrm{Tr}\check\tau_K 
  \langle\check{{\mathbf j}}({\mathbf{r}},{\mathbf{v_F}},E)
  \rangle_{{\mathbf{p_F}}},
\end{eqnarray}
where $\langle\rangle_{{\mathbf p_F}}$ denotes angular averaging of the
momentum direction and $\check\tau_K=\hat\sigma_3\bar\tau_3$.

Physically, we have in this way integrated out the \emph{fast} spatial
oscillation on a scale of the Fermi wave length $\lambda_F$, and the
remaining function $\check g$ is slowly varying on this scale. We have
also assumed that the selfenergy $\check\Sigma$ does not depend on the
momentum $p$. This has two important consequences.  First, the equation
is now much simpler and has the intuitive form of a transport equation
along classical trajectories. It is in fact easy to show, that
Eq.~(\ref{eq:eilenberger}) reproduces the well known Boltzmann equation for
normal transport.  Second, we have a price to pay.  Obviously the above
derivation fails in the vicinity of interfaces or singular points. This
means, that the boundary conditions have to be derived using the
underlying non-quasiclassical theory. Fortunately, this can be done
quite generally and one simply has to use effective boundary conditions.
Related to the problem of boundary conditions is the homogeneity of
Eq.~(\ref{eq:eilenberger}). In fact, it was shown by Shelankov
\cite{shelankov:85} that the condition ensuring the regularity for $x\to
x'$ in the general Green's function leads to a normalization condition
\begin{equation}
  \label{eq:normalization}
  \check g^2({\mathbf r},{\mathbf{v_F}},E)=\check1 .
\end{equation}
This condition is very important, since almost any further manipulation
of Eq.~(\ref{eq:eilenberger}) relies on it.

In the context of superconductivity the most important contribution to
the self energy is the pairpotential
\begin{equation}
  \label{eq:quasi-pairpotential}
  \hat\sigma^{R(A)}=-i\hat\Delta\;,\;
  \hat\Delta=-i\frac{\lambda}{4}\int dE 
  \langle\hat g^{K}_{\textrm{\tiny offdiag}}\rangle_{{\mathbf p}_F}\,,
\end{equation}
where 'offdiag' denotes that only the off-diagonal components in Nambu
space should be considered. $\lambda$ is the attractive BCS interaction
constant.

\subsection{The Dirty Limit -- Usadel Equation}
\label{sec:usadel}

An important simplification arises if the system is almost homogeneous in the
momentum direction. This is e.g. the case for diffusive systems, in which the
dominant term in the self-energy arises from impurity scattering. In
(self-consistent) Born approximation the impurity self-energy has the form
\begin{equation}
  \label{eq:impurity-selfenergy}
  \check\sigma({\mathbf r},E)=
  -\frac{i}{2\tau_{imp}} 
  \langle\check g({\mathbf r},{\mathbf p_F},E)\rangle_{{\mathbf p_F}}\,.
\end{equation}
In the limit $1/\tau_{imp}\gg E,\Delta,$ etc., the Green's functions will
be nearly isotropic and we make the Ansatz $\check g({\mathbf r},{\mathbf
p_F},E)=\check G({\mathbf r},E)+{\mathbf p_F} \check{{\mathbf G}}({\mathbf r},E)$.
Using Eq.~(\ref{eq:eilenberger}) together with the normalization condition we
obtain the so-called Usadel equation \cite{usadel}
\begin{equation}
  \label{eq:usadel-appendix}
  \frac{1}{e^2N_0}\nabla \check{{\mathbf j}}({\mathbf r},E) =
  \left[ -iE +\check\Delta,\check
  G({\mathbf r},E)\right]\,,
\end{equation}
where the generalized matrix current is now
\begin{equation}
  \label{eq:usadel-matrix-current}
  \check{\mathbf j}({\mathbf r},E)=\sigma \check G({\mathbf r},E)
  \nabla\check G({\mathbf r},E)
\end{equation}
The conductivity is given by the Einstein relation $\sigma=e^2N_0D$,
with the diffusion coefficient $D=v_F^2\tau_{imp}/3$. The current
density is
\begin{equation}
  \label{eq:current-usadel}
  {\mathbf{j}}({\mathbf{r}})=\frac{1}{4e}\int dE {\textrm{Tr}}
  \left[\check\tau_K\check{\mathbf{j}}({\mathbf r},E)\right]\,.
\end{equation}

\subsection{Boundary conditions}
\label{sec:bc}

Close to boundaries the quasiclassical equations are invalid and have to
be supplemented by boundary conditions. In the general case these
boundary conditions have been derived by Zaitsev \cite{zaitsev:84}. They
are rather complicated and we will not treat them here. In diffusive
systems a very concise form of the boundary conditions was obtained by
Nazarov\cite{yuli:99-supplat}. Under the assumption that two diffusive
pieces of metals are connected by a quantum scatterer, the matrix
current depends only on the ensemble of transmission eigenvalues and has
the form (\ref{eq:matrix-current}). The boundary condition is equivalent
to the conservation of matrix currents in the adjacent metals.

\subsection{Bulk Solutions}
\label{sec:bulk-solutions}

We summarize the necessary ingredients for the various circuits
treated in this review. As terminals we consider only normal metals or
superconductors. They are determined by external parameters like applied
potentials or temperature. The matrix structures are obtained from the
bulk solutions of the Eilenberger- or Usadel-equations. We give below
their form in the triangular Keldysh-matrix representation
(\ref{eq:keldysh-matrix}).

A normal metal at chemical potential $\mu$ and temperature $T_e$ is
described by a Green's function
\begin{equation}
  \label{eq:green-normal}
  \underline{G}_N(E)=\hat\sigma_3\bar\tau_3
  +(\bar\tau_1+i\bar\tau_2)
  \left(1-f(E)-f(-E)+\hat\sigma_3(f(-E)-f(E))\right)
\end{equation}
with the Fermi distribution $f(E)=(\exp((E-\mu)/k_BT_e)+1)^{-1}$.  

A superconducting terminal at chemical potential $\mu_S=0$ is described
by
\begin{equation}
  \label{eq:green-super}
  \underline{G}_S(E)=\frac12\left(\hat R+\hat A\right)
  +\frac12\left(\hat R-\hat A\right)
  \left(\bar\tau_3+(\bar\tau_1+i\bar\tau_2)
    \tanh\left(\frac{E}{2k_BT_e}\right)\right)\;,
\end{equation}
where the retarded and advanced functions are
\begin{equation}
  \hat R(\hat A) = 
  \left(
    \begin{array}[c]{cc}
      g_{R(A)} & f_{R(A)} \\
      f_{R(A)}^\dagger & - g_{R(A)}
    \end{array}
  \right)=
  \frac{(E\pm i\delta)\hat\sigma_3+i\hat\Delta}{
    \sqrt{\left(E\pm i\delta\right)^2-|\Delta|^2}}
  \,,
\end{equation}
where $\delta$ is a broadening parameter.
The gap matrix contains the dependence on the superconducting phase $\phi$
\begin{equation}
  \label{eq:gap-matrix}
  \hat\Delta=\left(
    \begin{array}[c]{cc}
      0 & |\Delta|e^{i\phi} \\
      |\Delta|e^{-i\phi} & 0 \\
    \end{array}
    \right)\,.
\end{equation}
In the limit of zero temperature and $|E|\ll\Delta$ the bulk
solutions simplify to
\begin{eqnarray}
  \label{eq:bulk-zerotemp}
  \underline{G}_N(E) & = & 
  \left\{
    \begin{array}[c]{lll}
      \hat\sigma_3\bar\tau_3
      -(\bar\tau_1+i\bar\tau_2)\textrm{sgn}(eV)&,&
      |E|<|eV|\\
      \hat\sigma_3\bar\tau_3
      -(\bar\tau_1+i\bar\tau_2)\hat\sigma_3\textrm{sgn}(E) &,&
      |E|>|eV|
    \end{array}
  \right.
  \\
  \underline{G}_S(\phi) & = & \hat\sigma_1 \cos(\phi) -\hat\sigma_2\sin(\phi)\,.
\end{eqnarray}

\section{Appendix: CGF for the Single Channel Contact}

\subsection{Super-Normal contact: coefficients $A_n$}

\label{sec:coefficients-sn}
We present the coefficients in the CGF (\ref{eq:cgf-sncontact}) for a
channel of transparency $T_n$. We assume the superconductor to be
in equilibrium and the normal metal at a chemical potential $\mu_N$. The
occupation factors are $f_{\pm}^N=f_0(\pm E-\mu_N)$ and
$f_{\pm}^S=f_0(\pm E)$, where $f_0$ is the Fermi-Dirac distribution. The
coefficients take the form
\begin{eqnarray}
  \label{eq:a1}
  A_{n,1} & = & T_n(1-T_n/2)\frac{2(g_R-g_A)}{(2-T_n(g_R-1))(2-T_n(g_A+1))}\\ 
  &&\quad\times\nonumber
  \left[f_+^N(1-f_+^S)+f_-^N(1-f_-^S)\right] 
  \\ &&\nonumber
  +2T^2_n\frac{1-f_Rf_A-g_Rg_A}{(2-T_n(g_R-1))(2-T(g_A+1))}
  \\ &&\quad\times\nonumber
  \Big[(f_+^N-f_-^N)(f_+^S-f_-^S)(1-(f_+^N-f_-^N)(f_+^S-f_-^S))
  \\ && \qquad\quad\nonumber
  +2 (f_+^S-f_-^S)^2(1-f_+^N)(1-f_-^N)\Big]\,,\\\label{eq:a2}
  A_{n,2} & = & \frac{T^2_n}{2} f_+^Nf_-^N\times\\\nonumber &&
  \frac{1+f_Rf_A-g_Rg_A-(f_+^S-f_-^S)^2(1-f_Rf_A-g_Rg_A)}{
    (2-T_n(g_R-1))(2-T_n(g_A+1))}\,.
\end{eqnarray}
The coefficients $A_{n,-q}$ can be obtained from Eq.~(\ref{eq:a1}) and
Eq.~(\ref{eq:a2}) by the substitution $f_+^{S(N)} \leftrightarrow
(1-f_-^{S(N)})$, i.e. interchanging electron-like and hole-like
quasiparticles.

\subsection{Super-Super contact: coefficients $A_n^S$}

\label{sec:coefficients-ss}

Introducing $q=(1-g_{\rm R}g_{\rm A})(1-h^2) +f_{\rm R}f_{\rm A}(1+h^2)$
the coefficients may be written as
\begin{eqnarray}
  \label{eq:coeffa2}
  A_{n,\pm 2}^S & = & \frac{T^2_n}{64}q^2,\\
  \label{eq:coeffa1}
  A_{n,\pm 1}^S & = &  \frac{T_n}{4}q-
  \frac{T^2_n}{16}q\left[q-4f_{\rm R}f_{\rm A}\sin^2\frac\phi2\right]
  \\\nonumber&&
  +\frac{T_n}{8}\left[(f_{\rm R}+f_{\rm A})h\cos\frac\phi2\mp 
    i(f_R-f_A)\sin\frac\phi2\right]^2,\\
  \label{eq:coeffq}
  Q_n & = & \left[1-T_nf_{\rm R}^2\sin^2\left(\frac{\phi}{2}\right) \right]
  \left[1-T_nf_{\rm A}^2\sin^2\left(\frac{\phi}{2}\right)\right].
\end{eqnarray}

\end{document}